\documentstyle[12pt]{article}
\def\be{\begin{equation}}
\def\fe{\end{equation}}

\def\ssum{ {\hbox{$\sum\,$}} }

\def\spose#1{\hbox to 0pt{#1\hss}}\def\lta{\mathrel{\spose{\lower 3pt\hbox
{$\mathchar"218$}}\raise 2.0pt\hbox{$\mathchar"13C$}}}  \def\gta{\mathrel
{\spose{\lower 3pt\hbox{$\mathchar"218$}}\raise 2.0pt\hbox{$\mathchar"13E$}}} 

\def\Libra{\spose {--} {\cal L}}

\begin{document}

\title{\bf Covariant analysis of Newtonian multi-fluid
models for neutron stars:  II Stress - energy tensors
and virial theorems}

\author { {\bf Brandon Carter \& Nicolas Chamel }\\
 \hskip 1 cm\\   \\Observatoire de
Paris, 92195 Meudon, France.}

\date{\it December, 2003}

\maketitle 

\vskip 1 cm
{\bf Abstract} The 4-dimensionally covariant approach to  multiconstituent 
Newtonian fluid dynamics presented in the preceding article of this series 
is developed by construction of the relevant 4-dimensional stress energy
tensor whose conservation in the non-dissipative variational case is
shown to be interpretable as a Noether identity of the Milne spacetime
structure. The formalism is illustrated by the application to homogeneously
expanding cosmological models, for which appropriately generalised local
Bernouilli constants are constructed. Another application is to the
Iordanski type generalisation of the Joukowski formula for the Magnus
force on a vortex. Finally, at a global level, a new (formally simpler but
more generally applicable) version of the ``virial theorem'' is obtained
for multiconsituent - neutron or other - fluid star models as a special
case within an extensive category of formulae whereby the time evolution
of variously weighted mass moment integrals is determined by corresponding
space integrals of stress tensor components, with the implication that
all such stress integrals must vanish for any stationary equilibrium
 configuration.

\vskip 1.6 cm

\bigskip
{\bf 1. Introduction}
\medskip

This article is the second of a series providing a systematic treatment of 
the essential dynamical properties of multifluid models (as needed for the 
description of a neutron superfluid moving with respect to a normal 
background) using a non-relativistic but  4-dimensionally covariant 
formalism in which the preferred Newtonian time gradient $t_\mu$ features
as a null eigenvector of the degenerate contravariant space metric tensor 
$\gamma^{\mu\nu}$. In the preceding aricle~\cite{CCI}, which will be 
referred to simply as (I), it was shown how, in the absence of dissipation,
the equations of motion are obtainable, in the conservative limit,
from a variation principle given, in a fixed background gravitational
field $\phi$, by a Lagrangian $\Lambda$ that is specified as a function 
of a set of conserved current 4-vectors $n_{_{\rm X}}^{\,\mu}$. Subject to
the condition that the variation of these currents be restricted to have an 
appropriate world line dependent form, the dynamical equations were derived 
as the condition of vanishing of a corresponding set of 4-force densities 
$f^{_{\rm X}}_{\,\nu}$ that were specified by the equation
(I-159)  in terms of the dynamical conjugates of the currents,
namely the set of 4-covectors $\pi^{_{\rm X}}_{\,\nu}$ representing
the effective energy-momentum per particle of the species labelled
by the suffix $X$.

The present article (II) describes the generalisation needed to allow for 
active self gravitational effects, which can be dealt with in the 
conservative case simply by adding an appropriate gravitational field 
contribution $\Lambda_{_{\rm grf}}$ to the action density. It will be 
shown how the associated stress-momentum- energy density 4-tensor will 
be subject to various (generalised Bernouilli and virial type) 
conservation laws ensuing from the invariance of the theory with respect 
not just to Galilean transformations but also to their accelerated, 
Milne type, generalisations, (which are important for cosmological 
applications). A formal derivation of the relevant Noether identities is 
provided in an appendix. A following article (III) will describe the 
generalisation needed to allow for non dissipative effects such as 
viscosity, resistivity, and drag on superfluid vortex lines. 

\bigskip
{\bf 2. Stress momentum energy tensor}
\medskip

Experience with the technically simpler relativistic case~\cite{Carter89} 
(and previous work on the Newtonian case~\cite{CarterKhalatnikov94}) 
suggests the utility at this stage of introducing the stress momentum 
energy density tensor
\be T^\mu_{\ \,\nu}=\ssum n_{_{\rm X}}^{\,\mu}\pi^{_{\rm X}}_{\,\nu}
+\Psi\delta^\mu_{\,\nu}\, ,\label{80}\fe
in which the covectors $\pi^{_{\rm X}}_{\,\nu}$ are the 4-momenta defined
by (I-152) and  $\Psi$ is the generalised pressure function defined by
(I-148). The utility of this (chemically covariant) tensor derives from
the property that its divergence can be seen to be given just by the sum 
\be f_\mu=\ssum f^{_{\rm X}}_{\,\mu} \, ,\label{80f}\fe
of the relevant non gravitational force densities -- which could be expected 
to cancel each other out, even in a non-conservative model, so long as it is 
isolated from external interactions -- plus the relevant gravitational force 
density which can always be expected to be present, by the simple formula
\be \nabla_{\!\mu} T^\mu_{\ \,\nu}=f_{\nu}-\rho\nabla_{\!\nu}\phi
\, ,\label{81}\fe
a relation that is shown in the appendix to be interpretable as a 
Noether identity. 

It is to be remarked that the 3-momentum density vector given by (I-146)
will be obtainable directly from (\ref{80}) as the mixed space and time 
projection specified by
\be {\mit \Pi}^\mu = t_\rho T^\rho_{\ \nu}\gamma^{\nu\mu}
\, .\label{81a}\fe
The gauge independent character of this 3-momentum density
${\mit \Pi}^\mu$, as manifested by its original specification (I-145)
is to be contrasted with the highly gauge dependent character of the 
corresponding energy density current, namely
\be U^\mu=-T^\mu_{\ \,\nu}e^\nu \, ,\label{81b}\fe
and in particular of its time component, the ordinary energy density,
which will be given by
\be U=U^\mu t_\mu =-\ssum n_{_{\rm X}}\pi^{_{\rm X}}_{\,\nu}e^\nu
-\Psi \, .\label{81c}\fe
The latter will evidently be decomposable in the form
\be U=U_{_{\rm mat}}+U_{_{\rm pot}}\, ,\label{81d}\fe
in which the potential energy contribution is just the opposite
of the corresponding Lagrangian contribution (I-94), i.e.
\be U_{_{\rm pot}}=\rho\phi=-\Lambda_{_{\rm pot}}\, ,\label{81e}\fe
while the material contribution will be given by
\be U_{_{\rm mat}}=-\ssum n_{_{\rm X}}\mu^{_{\rm X}}_{\,\nu}e^\nu
-\Psi \, .\label{81f}\fe

The complete stress energy momentum density tensor will evidently have a 
corresponding decomposition as the sum of a purely material part and a 
gravitational contribution in the form
\be T^\mu_{\ \,\nu}=T_{_{\!\rm mat}\,\nu}^{\ \ \mu}-\phi\rho^\mu
t_\nu \, ,\label{82}\fe
with
\be T_{_{\!\rm mat}\,\nu}^{\ \ \mu}=\ssum n_{_{\rm X}}^{\,\mu}
\mu^{_{\rm X}}_{\,\nu}+\Psi\delta^\mu_{\,\nu}\, ,\label{82a}\fe
in which $\Psi$ is the same as before, i.e. there is no need to distinguish 
between $\Psi$ and $\Psi_{_{\rm mat}}$, because its gravitational 
contribution cancels out so that it is directly expressible as
\be \Psi=\Lambda_{_{\rm mat}} -\ssum n_{_{\rm X}}^{\,\nu}
\mu^{_{\rm X}}_{\,\nu} \, ,\label{82b}\fe
and its general variation (I-151) simplifies as
\be \delta\Psi= -\ssum n_{_{\rm X}}^{\,\nu}\,\delta
\mu^{_{\rm X}}_{\,\nu} \, .\label{82c}\fe
Actually we can take the decomposition one step further by expressing the 
material contribution as the sum of a kinetic contribution and a purely 
internal (and therefore gauge invariant as well as chemically covariant) 
internal part in the form
\be T_{_{\!\rm mat}\,\nu}^{\ \ \mu}= T_{_{\!\rm kin}\,\nu}^{\ \, \mu}+
 T_{_{\!\rm int}\,\nu}^{\ \, \mu}\, ,\label{83}\fe
where the kinetic contribution is given simply by
\be T_{_{\!\rm kin}\,\nu}^{\ \, \mu}=\ssum n_{_{\rm X}}^{\,\mu}
p^{_{\rm X}}_{\,\nu}\, \label{83k}\fe
and the internal contribution is the Milne gauge invariant {\it pressure 
tensor}, $P^\mu_{\ \nu}$ say, which is given by
\be T_{_{\!\rm int}\,\nu}^{\ \,\mu}=P^\mu_{\ \nu}= \ssum n_{_{\rm X}}^{\,\mu}
\chi^{_{\rm X}}_{\,\nu}+\Psi\delta^\mu_{\,\nu}\, ,\label{83a}\fe
with $\Psi$ given in terms of quantities obtained just from
the internal contribution to the Lagrangian by the formula (I-148).
The corresponding decomposition of the material energy contribution
(\ref{81f}) will have the form
\be U_{_{\rm mat}}= U_{_{\rm kin}}+U_{_{\rm int}} \, .\label{83f}\fe
in which it can be seen, using (I-48), that we shall have
\be U_{_{\rm kin}}=\Lambda_{_{\rm kin}}={_1\over^2}\ssum
 \gamma^\nu_{\,\mu}n_{_{\rm X}}^{\,\mu}p^{_{\rm X}}_{\,\nu}\, ,
\hskip 1 cm U_{_{\rm int}}=\ssum \gamma^\nu_{\,\mu}n_{_{\rm X}}^{\,\mu}
\chi^{_{\rm X}}_{\,\nu}-\Lambda_{_{\rm int}}\, .\label{83h}\fe
In terms of the ether rest frame chemical potentials defined by
\be \chi^{_{\rm X}}=-\chi^{_{\rm X}}_{\,\nu}\, e^\nu\label{83e}\fe
it can be seen that we shall have
\be U_{_{\rm int}}=\ssum n_{_{\rm X}}\chi^{_{\rm X}}-\Psi
\, ,\hskip 1 cm \delta U_{_{\rm int}}=\ssum\chi^{_{\rm X}}\,\delta
n_{_{\rm X}} +\ssum \gamma^\nu_{\,\mu}n_{_{\rm X}}^{\,\mu}\,
\delta\chi^{_{\rm X}}_{\,\nu}\, .\label{83g}\fe

It follows from the identity (I-144) that the contraction of the 
pressure tensor with the degenerate space metric $\gamma^{\mu\nu}$ gives 
a result that is symmetric, i.e.
\be \gamma^{\rho[\mu}P^{\nu]}_{\ \,\rho}=0\, .\label{84}\fe
It can be seen that this entails a corresponding symmetry property for the 
space projected part of the complete stress momentum energy density, namely
\be \gamma^\rho{_{\,[\mu}\gamma_{\nu]\sigma}} T^\sigma_{\ \rho}=0
\, .\label {84b}\fe

Putting all the pieces together again, we see that this complete stress 
momentum energy density tensor will be expressible as
\be T^\mu_{\ \nu}=P^\mu_{\ \nu}+\ssum n_{_{\rm X}}^{\,\mu} \big(
p^{_{\rm X}}_{\,\nu}-m^{_{\rm X}}\phi\,  t_\nu\big)\, .\label{84a}\fe

\bigskip
{\bf 3. Action and stress energy for the gravitational field}
\medskip

Up to this point we have been treating the gravitational field just as 
a given background, but we can promote it to the status of an active 
dynamical field by taking $\phi$ to be an extra independent variable 
in the Lagrangian and adding in an extra gravitational field term  so
as to obtain a total action density given by
\be \Lambda_{_{\rm tot}}=\Lambda+\Lambda_{_{\rm grf}}\, ,\label{85}\fe
in which the gravitational field term is given by
\be \Lambda_{_{\rm grf}}= (8\pi {\rm G})^{-1}g^\mu
\nabla_{\!\mu}\phi \, ,\label{85a}\fe
where the graviational field is defined according to the potential $\phi$ as
\be g^\mu=-\gamma^{\mu\nu}\nabla_{\!\nu}\phi .\fe
Since the only other contribution involving $\phi$ is the potential 
energy term $\Lambda_{_{\rm pot}}$ given by (I-136), it can 
immediately be seen that the requirement of invariance of the total 
action with respect to localised perturbations of $\phi$ does indeed 
give the usual Poisson source equation, which, in the covariant formulation 
we are using here, will be given by 
\be \gamma^{\mu\nu}\nabla_{\!\mu}\nabla_{\!\nu}\phi=
4\pi{\rm G}\,\rho\, .\label{85b}\fe
This can be alternatively presented in a manner reminiscent of the
relativistic Einstein equation, using the notation of the Newton Cartan 
formalism described above, as
\be R_{\mu\nu}= 4\pi {\rm G}\, \rho\,t_\mu t_\nu 
\, ,\label{85c}\fe
where $R_{\mu\nu}$ is the Ricci type trace of the Newton Cartan 
curvature, as given by (I-31). 

For the reason discussed in the appendix, there will be an associated 
gravitational stress energy momentum density tensor given by the formula 
\be T_{_{\!\rm grf}\nu}^{\ \mu}= -(4\pi{\rm G})^{-1}
\big(g^\mu\nabla_{\!\nu}\phi-{_1\over^2}\delta^\mu_{\,\nu}
g^\rho\nabla_{\!\rho}\phi\big)\label{86}\fe
(which was misstyped in the previous version~\cite{CarterKhalatnikov94}
where the initial minus sign was omitted). 

The purely gravitational contribution (\ref{86}) can be added to the 
contribution (\ref{84a}) associated with the material source distribution 
to give a grand total
\be  T_{_{\!\rm tot}\nu}^{\ \mu} =T^\mu_{\ \,\nu}+
T_{_{\!\rm grf}\nu}^{\ \, \mu} \, ,\label{86a}\fe
that satisfies the identity
\be \nabla_{\!\mu}T_{_{\!\rm tot}\nu}^{\ \mu}
=f_{\nu}\, ,\label{86b}\fe
in which all that remains on the right is the sum (\ref{80f}) of the non 
gravitational force density contributions, if any, that are defined by 
the variational formula (I-159).

In the strictly conservative case for which the variational field equations 
\be f^{_{\rm X}}_{\,\nu}=0\, ,\label{86c}\fe
are satisfied, and much more generally in any system that is effectively 
isolated so that the separate contributions in (\ref{86c}) cancel out in 
the sum (\ref{80f}), leaving a vanishing total force density, $f_\mu=0$,
we shall simply be left with a total energy momentum conservation law of 
the form
\be \nabla_{\!\mu}T_{_{\!\rm tot}\nu}^{\ \mu}=0\, ,\label{86d}\fe
in which the  conserved tensor field (\ref{86a}) can be redecomposed in 
the form
\be T_{_{\!\rm tot}\nu}^{\ \mu}= T_{_{\!\rm mat}\nu}^{\ \mu}+
T_{_{\!\rm grt}\nu}^{\ \mu}\, ,\label{87}\fe
where the total gravitational contribution is given by
\be T_{_{\!\rm grt}\nu}^{\ \mu}= T_{_{\!\rm pot}\nu}^{\ \mu}+
T_{_{\!\rm grf}\nu}^{\ \mu}\, .\label{87a}\fe

It is to be noted that there will be a corresponding total energy density
\be U_{_{\rm tot}}=-t_\mu T_{_{\!\rm tot}\nu}^{\ \mu} e^\nu
=U+U_{_{\rm grf}}\label{87b}\fe
in which the contribution (\ref{86}) provides a gravitational field energy 
density that can be evaluated, using (\ref{81b}), as
\be U_{_{\rm grf}}=-t_\mu T_{_{\!\rm grf}\nu}^{\ \mu} e^\nu=-\Lambda_{_{\rm grf}}=
-{_1\over^2}\rho\phi -(8\pi{\rm G})^{-1} \nabla_{\!\mu}\big(
\phi g^\mu\big)\, .\label{88}\fe
It can thereby be seen from (\ref{81d}) that the total energy density will be 
expressible in the form
\be U_{_{\rm tot}}=U_{_{\rm mat}} +U_{_{\rm grt}}\label{87c}\fe
in which the total gravitational energy contribution
\be U_{_{\rm grt}}= U_{_{\rm pot}}+ U_{_{\rm grf}}\label{88a}\fe
will be given by
\be U_{_{\rm grt}}=-t_\mu T_{_{\!\rm grt}\nu}^{\ \mu} e^\nu
={_1\over^2}\rho\phi -(8\pi{\rm G})^{-1} \nabla_{\!\mu}\big(
\phi g^\mu\big)\, .\label{88b}\fe
It is to be observed that the final divergence term in (\ref{88}) and
(\ref{88b}) will cancel itself out when integrated over the volume 
surrounding a confined source provided that (as will always be possible in 
such a case) the gauge is chosen in such a way that the large distance limit 
of $\phi$ is zero. The remaining term in (\ref{88}) will then half cancel the 
corresponding potential energy contribution given by (\ref{81e}), leaving 
a remaining term in (\ref{88b}) that is only half of $U_{_{\rm pot}}$. The 
effect of this semi-cancellation is to provide a total in which the 
gravitational binding between each pair of particles is (as it should be) 
counted only once, not twice.
 
Another point to be noted is that there will be no corresponding 
gravitational field contribution to the space-like 3-momentum density 
(\ref{81a}), i.e. we shall have
\be t_\rho T^{\ \rho}_{_{\!\rm grf} \nu}\gamma^{\nu\mu}=0
\, .\label{89}\fe
It can thus be seen from (\ref{81b}) that (\ref{81a}) will be replaceable by 
the equivalent specification
\be {\mit \Pi}^\mu = t_\rho T^{\ \rho}_{_{\rm tot} \nu}\gamma^{\nu\mu}
\, .\label{89a}\fe

\bigskip
{\bf 4. Space time symmetries and homogeneous cosmological models.}
\medskip

Let us now consider cases in which all or part of the system under study is 
invariant with respect to the action of one or several symmetry generators, 
$k_{\rm a}^\mu$,  ${\rm a}=0,1, ... $. (For example in an application to 
a neutron star model, $k_{0}^{\,\mu}$ might be a stationarity generator, 
i.e. the generator of a one parameter time translation group (in which 
case it would specify a natural choice of ether vector by the 
identification $e^\mu= k_{0}^{\,\mu}$) and $k_{1}^{\,\mu}$ might be
an axisymmetry generator, i.e. the generator of a one parameter group 
of rotations about some preferred symmetry axis. This means that all 
physically well defined fields must be invariant under the action of the 
corresponding Lie differentiation operators which we shall designate by 
the short hand notation $\Libra_{\rm a}\equiv \vec k_{\rm a}\Libra$,
or simply $ \Libra=\vec k\Libra $ when we are only considering a
single generator so that no ambiguity arises. It is to be emphasised 
however that physical invariance does not necessarily require 
vanishing of the Lie derivative of a field $q$ that is gauge dependent, 
but only that its Lie derivative $\Libra q$ should be cancellable by some 
infinitesimal gauge transformation of the kind denoted by $\breve {\rm d} q$ 
in the discussion at the end of Section 6 of \cite{CCI}. More specifically, all that is
required is that for each of the relevant Lie differentiation  operators 
$\Libra_{\rm a}$ there should be a corresponding  infinitesimal gauge 
transformation operator $\breve {\rm d}_{\rm  a}$ such that any relevant 
field satisfies the condition
\be \Libra_{\rm a} q+\breve {\rm d}_{\rm a} q=0\, .\label{94}\fe 
The occurrence of symmetries of this more general kind, involving 
non-vanishing infinitesimal gauge transformations, is exemplified by the
noteworthy case of the Milne type~\cite{Bonnor57} homogeneous 
cosmological models that are described below.

A minimal requirement for any symmetry generator $k_{\rm a}^{\,\mu}$ is
that its action should preserve the Milne structure of Newtonian space 
time. For the basic Coriolis structure fields $t_\mu$ and $\gamma^{\mu\nu}$ 
the question of whether there is a corresponding gauge transformation operator
$\breve{\rm d}_{\rm a}$ does not arise, since they are gauge independent, 
so all that is required is the vanishing of their Lie derivatives, conditions 
that reduce by (I-17) to the form
\be t_\nu\nabla_{\!\mu}k_{\rm a}^{\,\nu}=0\, ,\hskip 1 cm \gamma^{\rho(\mu}
\nabla_{\!\rho} k_{\rm a}^{\,\nu)}=0\, ,\label{94a}\fe
from which the space projected  and complete space time divergence
conditions
\be \gamma^\mu_{\,\nu} \nabla_{\!\mu} k_{\rm a}^{\,\nu}=0\, ,\hskip 1 cm
\nabla_{\!\nu} k_{\rm a}^{\,\nu}=0\, ,\label{94b}\fe 
are obtained as corollaries. In a particular Aristotelian frame with
respect to an ordinary orthonormal system of coordinates $\{t,X^i\}$, the 
 conditions (\ref{94a}) will be expressible as 
\be \nabla_{\!_0} 
k_{\rm a}^{\,_0}=0\, ,\hskip 1 cm \nabla_{\!i}k_{\rm a}^{\,_0}=0
\, ,\hskip 1 cm  \nabla^{\,(i}k_{\rm a}{^{j)}} =0\, .\label{94f}\fe 

In the context of general relativity the well known condition for
$k_{\rm a}^{\,\mu}$ to be a symmetry generator of the spacetime structure is 
just the well known Killing equation $\nabla^{(\mu}k_{\rm a}{^{\nu)}}=0$, 
which guarantees the invariance of the spacetime metric $g_{\mu\nu}$ and 
hence also of the derived Riemannian connection. However in the Newtonian 
case, since the connection is not simply derivable from the structure fields 
$t_\mu$ and $\gamma^{\mu\nu}$, the corresponding first order differential 
conditions (\ref{94a}) will not by themselves be sufficient to qualify 
$k_{\rm a}^{\mu}$ as a space time symmetry generator. In order for 
$k_{\rm a}^{\,\mu}$ to qualify as a Newtonian space time symmetry generator 
its action must also preserve the flat connection 
$\Gamma_{\!\mu\ \rho}^{\ \nu}$. Since the latter is subject to the gauge 
dependence condition (I-75) its Lie derivative will not have to vanish 
absolutely, but only modulo the action of some infinitesimal gauge 
transformation, which will be characterised by some (infinitesimal) boost
potential $\beta_{\rm a}$, with a derived time (but not space) dependent 
boost vector $b^\mu$ and a corresponding acceleration vector $a^\mu$, that 
are given, as in (I-15) by 
\be b_{\rm a}^{\mu}=\gamma^{\mu\nu}\nabla_{\!\nu}\beta_{\rm a}\, ,\hskip 
1 cm \gamma^{\mu\nu}\nabla_{\!\nu} b_{\rm a}^{\,\rho}=0\, .\label{90b}\fe
and
\be a_{\rm a}^{\,\mu}=e^\nu\nabla_{\!\nu}b_{\rm a}^{\,\mu}=
\gamma^{\mu\nu}\nabla_{\!\nu}\alpha_{\rm a} \, ,
\hskip 1 cm \alpha_{\rm a}= e^\nu\nabla_{\!\nu}\beta_{\rm a}
 \, .\label{90a}\fe

Modulo a choice of rest frame at some arbitrary reference event, the 
specification, according to (I-16), of an ether frame field  $e^\mu$ 
is equivalent to the specification of a corresponding connection 
$\Gamma_{\!\mu\ \rho}^{\ \nu}$, namely the one with respect to which 
the vector field $e^\mu$ is mapped onto itself by parallel transport. The 
preservation of this connection by the action generated by 
$k_{\rm a}^{\,\mu}$ will therefore be ensured simply by the requirement 
that it should preserve the ether vector. Substituting this vector $e^\mu$ 
in place of $q$ in the general purpose preservation condition  (\ref{94}), 
one sees from the transformation rule (I-74) and the parallel transport 
property (I-16) that the requirement of preservation of the ether 
vector is equivalent to the condition that the relevant boost transformation
$b_{\rm a}^{\,\mu}$ should be given simply by
\be b_{\rm a}^{\,\mu} =e^\nu\nabla_{\!\nu}k_{\rm a}^{\, \mu}
\, .\label{93}\fe
It is evident that this will conveniently vanish, so that there will be no 
need to bother about allowance for the gauge adjustment, in cases for 
which the symmetry generator $k_{\rm a}^{\,\mu}$ is time independent, as
will be the case for the symmetries that are most relevant in applications
to rotating star models, namely stationarity and axisymmetry. On the other 
hand the gauge adjustment will have an indispensible role in the kind of 
time dependent translation symmetries that are relevant in homogeneous self 
gravitating (cosmological type) configurations such as will be described 
immediately below. In all cases, it can be seen from (\ref{94a}) that the 
formula (\ref{93}) can be used to express the gradient of the symmetry 
generator as the sum of orthogonally projected and mixed components in the 
form
\be \nabla_{\!\mu}k_{\rm a}^{\, \nu}= \gamma_\mu^{\,\rho}
\gamma_\sigma^{\,\nu}\nabla_{\!\rho}k_{\rm a}^{\, \sigma}+t_\mu\,
b_{\rm a}^{\, \nu}\, .\label{93b}\fe

In terms of the boost acceleration obtained from (\ref{93}) using 
(\ref{90a}), the invariance condition for the connection
$\Gamma_{\!\mu\ \rho}^{\ \nu}$ will be given according to the general 
principle (\ref{94}) by the formula
\be \nabla_{\!\mu}\nabla_{\!\nu} k_{\rm a}^{\,\rho}=t_\mu t_\nu 
a_{\rm a}^{\,\rho}\, ,\label{94c}\fe
whose derivation is based on the use of the simple flat space special 
case of the Yano formula~\cite{Yano55} for the Lie derivative of a 
symmetric connection. 

The general version of the Yano formula, including allowance for 
curvature, is needed for the application to the Newton-Cartan connection
(I-25) whose Lie derivative will have the form given by
\be \Libra_{\rm a}\,\omega_{\mu\ \nu}^{\ \, \rho}=D_\mu D_\nu k^\rho
+R_{\sigma\mu\ \nu}^{\ \ \, \rho} k^\sigma\, .\label{95}\fe
In the Newtonian case (unlike the relativistic case) the conditions
for invariance of the relevant (Milne) space time structure,
namely the conditions (\ref{94a}) and (\ref{94c}), are not sufficient
to ensure invariance of the gravitational field, as embodied in
the independent gauge invariant connection  
$\omega_{\mu\ \nu}^{\ \, \rho}$, or equivalently in the gauge dependent
field $g^\mu$.  In view of its gauge independence, the invariance
under the action of $k_{\rm a}^{\,\mu}$ of the Newton Cartan 
connection requires simply that its Lie derivative, as given above,
should vanish,
\be \Libra_{\rm a}\,\omega_{\mu\ \rho}^{\ \, \nu}=0\, .\label{95a}\fe
Subject to the Milne structure invariance conditions (\ref{94a}) and
(\ref{94c}), it can be seen from the formula (\ref{95}) 
(using the expression (I-29) for the curvature) that the
supplementary condition (\ref{95a}) for invariance of the gravitational
field reduces to the condition obtained by application of the general
requirement (\ref{94}) to the gauge dependent gravitational
field vector $g^\mu$. It evidently follows from (I-23) that
this invariance requirement will take the form
 \be \Libra_{\rm a}\, g^\mu= a_{\rm a}^{\,\mu}\, ,\label{90}\fe
in which the Lie derivative of the field will of course be given
by the well known commutator formula
\be \Libra_{\rm a}\, g^\mu= k_{\rm a}^{\,\nu}\nabla_{\!\nu}g^\mu -
g^\nu\nabla_{\!\nu}k_{\rm a}^{\,\mu}\, .\label{95b}\fe
It similarly follows from (I-24) that at the more highly gauge
dependent level of the gravitational potential $\phi$, the corresponding 
field invariance condition will be expressible simply as 
\be k_{\rm a}^{\,\nu}\nabla_{\!\nu}\phi=-\alpha_{\rm a}\, ,\label{95c}\fe
where $\alpha_{\rm a}$ is the acceleration potential given by 
(\ref{90a}).

The prototype illustration of symmetries that are non-trivial, in the 
sense of requiring a non zero gauge transformation according to (\ref{93}), 
is provided by the case of {\it homogeneous isotropic} cosmological type 
models, with expansion described in terms of a comoving radial scale factor 
$\sigma$ say, and 3-velocity $v^i$ given in terms of Cartesian space
coordinates $X^i$ by $v^i= H X^i$ where the (time dependent but spacially 
uniform) Hubble parameter is given by $H=\dot\sigma/\sigma$ (using a dot for 
ordinary time derivation). Since the pressure is postulated to be uniform it 
has no effect on the motion, so the fluid particles will be effectively in 
free fall with respect to the background gravitational field, which will be 
given by $\phi={2\over3}\pi G\rho\gamma_{ij} X^iX^j$.

One of the things that delayed the discovery of these configurations is 
their lack of {\it stationarity}, i.e. their essentially time dependent
nature. Their matter distributions, and in particular the density $\rho$
are {\it not}  invariant with respect to the action of the ordinary time 
translation symmetry generator $k_{0}^{\,\mu}$ that is definable in terms 
of the flat Aristotelian coordinates $\{t,X^i\}$ used above by
\be k_{0}^{\ _0}=1\, ,\hskip 1 cm k_{0}^{\ i}= 0
\, ,\label{99c}\fe
which means that it is simply identifiable with the relevant Aristotelian 
ether vector, 
\be k_{0}^{\,\mu}=e^\mu \, .\label{99e}\fe
The misguided presumption that the cosmological model would have to be 
stationary was one of the psychological obstacles that, prior to its 
demolition by Friedmann's theoretical insight (and Hubble's observations) 
had prevented everyone (including Einstein) from seriously attempting to 
carry out the non trivial but (compared with the development of general 
relativity) not so difficult technical step that was finally taken by Milne.

The stationarity generator (\ref{99e}) illustrates a distinction that does 
not arise in General relativity theory, where the property of being a Killing 
vector, i.e. a solution of the equation $\nabla^{(\mu} k_{\rm a}^{\,\nu)}=0$, 
ensures the invariance not only of the background metric $g_{\mu\nu}$ but 
also, automatically, of the gravitational field as embodied by the 
associated connection. In a Newtonian context it is necessary to make a
distinction between what may be termed weak Killing vectors, or 
{\it Killing-Milne vectors}, meaning those that just satisfy the conditions 
(\ref{94a}) and (\ref{94c}) for invariance of the background spacetime 
structure, and what may be termed strong Killing vectors or {\it  
Killing-Cartan vectors}, meaning those that not only satisfy the conditions
(\ref{94a}) and (\ref{94c}) for preservation of the Milne structure but also
the supplementary condition (\ref{90}) for invariance of the gravitational
field and hence of the complete Newton Cartan structure. Regardless of the
matter distribution, the stationarity generator $k_{0}^{\,\mu}$ given by the
specification (\ref{99c}) will always be a Killing-Milne vector, since it
obviously satisfies (\ref{94a}) and also (with vanishing boost acceleration)
(\ref{94c}). However in the particular  application described above
$k_{0}^{\,\mu}$ is not a Killing vector in the strong sense because it does
not satisfy (\ref{90}).

Although they are not stationary, the cosmological configurations described
above are obviously {\it isotropic} in the sense of being invariant (in this
case without any need for an associated boost transformation) under the
action of the set of ordinary angular symmetry generators
$k_{\rm j}^{\,\mu}$ that are definable for values ${\rm j}=1,2,3$ of the
index ${\rm a}$ as follows. For a rotation about an axis specified by a
space unit vector $\nu_{\rm j}^{\,i}$, with components given, with respect
to the flat Aristotelian coordinate system we have been using, simply by
\be \nu_{\rm j}^{\,i}=\delta_{\rm j}^{\,i} \, ,\label{99f}\fe
the corresponding angular symmetry generator will be given by the
prescription
\be k_{\rm j}^{\, _0}=0\, ,\hskip 1 cm k_{\rm j}^{\, i}= Y^i_{\ k}
\nu_{\rm j}^k \, ,\label{99a}\fe
in terms \cite{CarterMcLenaghan79} of the flat space Killing-Yano 2-form,
whose (Cartesian) components are given by
\be Y_{ij} =\varepsilon_{ijk}X^k \, ,\label{99b}\fe
and which is characterisable, even with respect to non Cartesian
coordinates, by the Killing-Yano equation
\be \nabla_{\!(i}Y_{j)k}=0\, .\label{99g}\fe

Whatever the matter distribution may be, it is evident that, like the
stationarity generator $k_{0}^{\,\mu}$, these axisymmetry generators
$k_{\rm j}^{\,\mu}$ for ${\rm j}=1,2,3$ will always satisfy (\ref{94a}) and
(again with vanishing boost acceleration) also (\ref{94c}), so that they
will always qualify as Killing vectors in the weak sense. In the particular
example of the cosmological application described above it is also evident
that (in this case unlike the stationarity generator $k_{0}^{\,\mu}$) the
rotation generators will satisfy the further condition (\ref{90}) that
qualifies them for description as Killing-Cartan vectors.

We now come to the most essential, though not so obvious, symmetry feature
of the cosmological configurations described above, which is that as well as
being isotropic they are also {\it homogeneous} (with respect to space but
not time) in the sense of being characterised by a set of time dependent
space translation symmetries with generators to which we shall attribute
negative label values, a=-1,-2,-3 (since the positive label values
a=1,2,3 have already been used up in the specification (\ref{99a}) of the
rotation symmetry generators).
With respect to the same ordinary Aristotelian system of orthonormal
coordinates $\{ t, X^i\}$ as above, these space translation symmetry
generators will be given by the specification
\be k_{\rm -j}^{\ _0}=0\, ,\hskip 1 cm k_{\rm -j}^{\ i}=
\sigma\delta_{\rm j}^i \, ,\label{99}\fe
for ${\rm j} =1,2,3$. What delayed the discovery of such symmetries (even
after Friedmann had overcome the psychological barrier of the stationarity
presumption)  until Milne saw how to exploit the gauge invariance, is that
(in order for the generators (\ref{99}) to qualify as Killing vectors in
both the weak and strong sense described above) the effect of their action
needs to be cancelled by the effect of corresponding (non-linearly) time
dependent (so non Galilean) transformations characterised by (spacially
uniform) boost vectors $b_{\rm -j}^{\ i}$ and boost potentials
$\beta_{\rm -j}$ given by the formulae
\be b_{\rm -j}^{\ i}=\dot\sigma \delta_{\rm j}^i\, ,\hskip 1 cm
\beta_{\rm -j}=\dot\sigma\gamma_{{\rm j}i}X^i\, .\label{99h}\fe
In particular the velocity field can immediately be seen to satisfy the
invariance condition $\Libra_{\rm -a}\, v^i=b_{\rm -a}^{\,i}$, while the
gravitational field will satisfy the corresponding invariance conditions
\be \Libra_{\rm -j}\,g^i=a_{\rm -j}^{\ i}\, ,\hskip 1 cm
\Libra_{\rm -j}\,\phi =-\alpha_{\rm -j},\label{99i}\fe
(so that the translation generators $k_{\rm -j}^{\ \mu}$ qualify as
Killing vectors in the strong sense) where
\be a_{\rm -j}^{\ i}=\ddot\sigma\delta_{\rm j}^{\,i}\, ,\hskip 1 cm
\alpha_{\rm -j}=\ddot\sigma\gamma_{{\rm j}i}X^i\, .\label{99j}\fe

By contraction with the stress momentum energy density tensor
$T^\mu_{\ \nu}$, we can use the symmetry generators $k_{\rm a}^{\,\mu}$
to construct corresponding generalised momentum currents
\be {\cal P}_{\!\rm a}^{\,\mu}=T^\mu_{\ \nu} k_{\rm a}^{\,\nu}
\, ,\label{100}\fe
of which, for example ${\cal P}_{\! 0}^{\,\mu}$ will be interpretable as
the negative of an ordinary energy current,  ${\cal P}_{\! 1}^{\,\mu}$ as
a current of ordinary angular momentum (about the $X^{_1}$ axis in the
Aristotelian coordinate system used above) while ${\cal P}_{\! -1}^{\ \mu}$
will be interpretable as a current transporting a kind of generalised linear
momentum (in the direction of the $X^{_1}$ axis). The latter would reduce to
a current of linear momentum of the ordinary kind in the non expanding (weak
gravitational coupling) limit characterised by a time independent value of
the factor $\sigma$ in the specification (\ref{99}).

Using the Killing vector conditions (\ref{94a}) in the decomposition
(\ref{93b}), it can be seen to follow from the stress momentum energy tensor symmetry
property (\ref{84b}), that we shall have
\be   T^\mu_{\ \nu}\nabla_{\!\mu} k_{\rm a}^{\,\nu}=T^\mu_{\ \nu}
t_\mu b_{\rm a}^{\, \nu} \, ,\label{101}\fe
where  $b_{\rm a}^{\, \mu}$ is the boost vector given by (\ref{93}).
As a corollary of this useful lemma, it evidently follows from the tensor
divergence formula (\ref{81}) that the corresponding generalised momentum
densities will have scalar divergences given by
\be \nabla_{\!\nu} {\cal P}_{\rm a}^{\,\nu} =k_{\rm a}^{\,\nu}\big(f_\nu
-\rho\nabla_{\!\nu}\phi\big)+{\mit \Pi}^\nu\nabla_{\!\nu}
\beta_{\rm a} \, .\label{102}\fe
On the right of this formula, the first term (involving the total
external force $f_\nu$ if any) is of a familiar kind, representing the
ordinary rate of energy loss (per unit volume) in the case of the
stationarity generator $k_{0}^{\,\mu}$ and representing torque density
(about the $X^{_1}$ axis) in the case of the axisymmetry generator
$k_1^{\,\mu}$, and in these cases the last term will be absent. This last
term (involving the gradient of the boost potential $\beta_{\rm a}$) will
only be needed in cases for which the symmetry generator
$k_{\rm a}^{\,\mu}$ is time dependent. In the case of an isolated system,
as characterised by a vanishing total (external) force density, $f_\mu=0$,
and by a total mass current $\rho^\mu$ that satisfies the Newtonian mass
conservation law (I-108), it can be seen that (\ref{102}) will be
expressible in the form
\be \nabla_{\!\nu} \big({\cal P}_{\rm a}^{\,\nu}-\beta_{\rm a}\rho^\nu\big)
=-\rho\big(\alpha_{\rm a}+k_{\rm a}^{\,\nu}\nabla_{\!\nu}\phi\big)
\, ,\label{102a}\fe
in which the right hand side will drop out altogether if the gravitational
potential $\phi$ satisfies the condition (\ref{95c}) of invariance under
the symmetry action generated by $k_{\rm a}^{\,\nu}$, a condition for which
it is necessary that  $k_{\rm a}^{\,\nu}$ should be not just a Killing-Milne
vector, as has been assumed in the derivation of (\ref{102}), but more
particularly a Killing-Cartan vector in the sense discussed above, meaning
that its action preserves the gravitational field as well as the background
spacetime structure.

Independently of this latter (field invariance) restriction, but subject to
the requirement that the boost contribution should be absent, an isolated
system will always be characterised by a conservation law for a
corresponding, suitably adjusted, total generalised momentum vector.
To start with, we define the (unadjusted) total momentum vector in the
obvious way, simply replacing $T^\mu_{\ \nu}$ in (\ref{100}) by the total
stress energy tensor (\ref{86a}) which gives
\be {\cal P}_{\!\rm a_{\rm tot}}^{\ \ \mu}=T_{_{\!\rm tot}\,\nu}^{\ \, \mu}
k_{\rm a}^{\,\nu}={\cal P}_{\rm a}^{\,\mu}+T_{_{\!\rm grf}\,\nu}^{\ \, \mu}
k_{\rm a}^{\,\nu}\, .\label{103a}\fe
In the case of an isolated system, as characterised by a vanishing total
(external) force density, $f_\mu=0$, and by a total mass current
$\rho^\mu$ that satisfies the Newtonian mass conservation law (I-108),
it can be seen that the analogue of (\ref{102}) for this total will
reduce to a  generalised total momentum divergence law of the form
\be \nabla_{\!\nu} \big({\cal P}_{\!\rm a_{\rm tot}}^{\ \ \nu}-
\beta_{\rm a}\rho^\nu\big)
=-\rho\alpha_{\rm a}, .\label{103}\fe
Using the Poisson source equation (\ref{85b}) and the identity
$\alpha_{\rm a}\nabla_{\!\nu}g^\nu=\nabla_{\!\nu}(\alpha_{\rm a}
g^\mu+\phi a_{\rm a}^{\,\nu})$ the terms can finally be regrouped in
a single divergence on the left in the form
\be \nabla_{\!\nu}{\cal P}_{\!\rm a_{\rm aug}}^{\ \ \nu}
=0\, ,\label{103b}\fe
expressing conservation of a suitably augmented total generalised
momentum current that is given by the prescription
\be{\cal P}_{\!\rm a_{\rm aug}}^{\ \ \nu} =
{\cal P}_{\!\rm a_{\rm tot}}^{\ \ \nu}-
\beta_{\rm a}\rho^\nu -(4\pi{\rm G})^{-1}
(\alpha_{\rm a}g^\mu+\phi a_{\rm a}^{\,\nu})\, .\label{103c}\fe
The augmentation contributed by the second, third, and fourth terms will
actually be needed only in cases for which $k_{\rm a}^{\,\mu}$ is time
dependent (as in the cosmological example characterised by (\ref{99}),
(\ref{99h}) and (\ref{99j}) as described above). The first term is all that
remains, i.e. the unadjusted total ${\cal P}_{\!\rm a_{\rm tot}}^{\ \ \nu}$
will be conserved as it stands, in the more familiar cases in which
$k_{\rm a}^{\,\mu}$ is a generator of ordinary time translations (in which
case the conserved vector will be proportional to energy flux) or rotations
(for which it will be interpretable as angular momentum flux). It is to be
emphasised that the validity of the generalised total momentum conservation
law (\ref{103b}) follows just from the postulate that $k_{\rm a}^{\,\mu}$ is
a Killing-Milne vector in the weak sense discussed above (meaning just that
its action preserves the background spacetime structure), so that (unlike
the results to be derived in the next section) it will be valid regardless
of whether or not the relevant material field configurations have any
corresponding symmetry property.

\bigskip
{\bf 5. Fluid symmetries and generalised Bernoulli theorems}
\medskip

In contexts for which the local gravitational coupling can be considered to
be sufficiently weak, it can be of interest to consider configurations
of the generic kind governed by (\ref{102}) in which a material medium,
such as the multiconstituent fluid dealt with here, is not necessarily
subject to the symmetries of the background space time structure and the
background gravitational field.

The purpose of the present section is to show how much stronger
conclusions can be drawn if the medium is itself invariant under the action
of a Killing Cartan vector $k_{\rm a}^{\,\mu}$ of the kind
characterised by the space time symmetry conditions (\ref{94a}), (\ref{94b})
and the gravitational symmetry condition (\ref{90}) discussed above.

In the case of a multiconstituent fluid, it can be seen from  the formula
(I-77) for $\breve {\rm d} \pi^{_{\rm X}}_{\,\mu}$ that in terms of the
boost potential, the corresponding symmetry conditions on the momentum
covectors will be given, according to the general principle (\ref{94}), by
\be \Libra_{\rm a}\,\pi^{_{\rm X}}_{\,\nu}= m^{_{\rm X}}
\nabla_{\!\nu}\beta_{\rm a}\, .\label{90d}\fe

Since, by its definition, the Lie derivative of the momentum covector will
be given by
\be \Libra_{\rm a} \pi^{_{\rm X}}_{\,\mu}=2 k_{\rm a}^{\,\nu}
\nabla_{\![\nu}\pi^{_{\rm X}}{_{\!\mu]}} +\nabla_{\!\mu}\big(
k_{\rm a}^{\,\nu}\pi^{_{\rm X}}{_{\!\nu}}\big)\, , \label{91}\fe
we see from the definition (I-161) of the vorticity that the symmetry
condition on the momentum will be expressible in the form
\be \varpi^{_{\rm X}}_{\,\mu\nu}k_{\rm a}^{\,\nu}=\nabla_{\!\mu}
{\cal B}^{_{\rm X}}_{\rm a}\, ,\label{91b}\fe
in which ${\cal B}^{_{\rm X}}_{\rm a}$ is a generalised momentum component
that is defined by the formula
\be  {\cal B}^{_{\rm X}}_{\rm a}=
k_{\rm a}^{\,\nu}\pi^{_{\rm X}}_{\,\nu}-m^{_{\rm X}}\beta_{\rm a}
\, ,\label{91a}\fe
and that can be identified as a generalisation of the historic Bernoulli
scalar.

It is immediately obvious from (\ref{91b}) that we obtain a strong
generalisation of the Bernoulli theorem to effect that if the flow of some
particular constituent is irrotational, then the corresponding Bernoulli
scalar will be uniform throughout, in the sense of being independent
of both space and time,
\be \varpi^{_{\rm X}}_{\,\mu\nu}= 0 \hskip 0.5 cm \Rightarrow \hskip
0.5 cm \nabla_{\!\nu}{\cal B}^{_{\rm X}}_{\rm a}= 0\, .\label{91d}\fe

In the case (of the kind originally considered by Bernoulli) of ordinary
time translation symmetry, as obtained by setting a=0 with the specification
(\ref{99c}), the corresponding value,
\be {\cal B}^{_{\rm X}}_{0}=\pi^{_{\rm X}}_{\,\nu} e^\nu\, ,\fe
will be interpretable as the negative of an effective energy per particle. Similarly
in the case of an ordinary rotation symmetry, as obtained, for example by
setting a=1 (for a rotation about the $X^{_1}$ axis) according to the
specification (\ref{99a}) the corresponding value,
\be{\cal B}^{_{\rm X}}_{1} =\pi^{_{\rm X}}_{\,\nu}k_{1}^\nu
\, ,\label{105}\fe
will be interpretable as an effective angular momentum per particle. In the
case of a boosted translation symmetry along for example the $X^{_1}$ axis,
as obtained by setting $a=-1$ in the specification
(\ref{99}), we obtain a generalised Bernoulli scalar
\be {\cal B}^{_{\rm X}}_{-1}= k_{-1}^{\ \nu}\pi^{_{\rm X}}_{\,\nu}
-m^{_{\rm X}}\beta_{-1} \, ,\label{106}\fe
of a new kind, whose interpretation is not so elementary, due to the
involvement of the boost potential $\beta_{-1}$.

In the general case, involving rotation or even transfer of matter between
different constituents, it can be seen by combining (\ref{91b}) with the
general force density formula (I-164) that the gradient of the
relevant generalised Bernoulli scalar along the flow lines will be given by
\be n_{_{\rm X}}^{\,\nu}\nabla_{\!\nu} {\cal B}^{_{\rm X}}_{\rm a}
=k_{\rm a}^{\,\nu}f^{_{\rm X}}_{\,\nu}+k_{\rm a}^{\,\nu}
\pi^{_{\rm X}}_{\,\nu} {\cal D}_{_{\rm X}} \, .\label{107}\fe
We thus obtain a generalised weak Bernoulli theorem to the effect that in
the conservative case of vanishing force density, the Bernoulli scalar will
be constant along each separate flow line,
\be f^{_{\rm X}}_{\,\nu}=0\, ,\hskip 0.5 cm {\cal D}_{_{\rm X}}=0 \hskip
0.5 cm \Rightarrow \hskip 0.5 cm n_{_{\rm X}}^{\,\nu}\nabla_{\!\nu}
{\cal B}^{_{\rm X}}_{\rm a}=0 \, .\label{108}\fe

A useful alternative presentation of this generalised Bernoulli theorem can
be obtained by introducing generalised momentum current densities that are
defined for each separate constituent by the prescription
\be {\cal P}^{_{\rm X}\,\mu}_{\rm a}={\cal B}^{_{\rm X}}_{\rm a}
n_{_{\rm X}}^{\,\mu}\, .\label{109}\fe
The theorem (\ref{107}) can thereby be reformulated as a divergence law
of the form
\be\nabla_{\!\nu} {\cal P}^{_{\rm X}\,\nu}_{\rm a}=k_{\rm a}^{\,\nu}
f^{_{\rm X}}_{\,\nu}+\beta_{\rm a}m^{_{\rm X}}{\cal D}_{_{\rm X}}
\, .\label{111}\fe

It can be seen that the generalised total (non gravitational) momentum
density (\ref{100}) is related to the sum of the separate constituent
contributions by
\be {\cal P}_{\rm a}^{\,\mu}-\beta_{\rm a}\rho^\mu
=\ssum {\cal P}^{_{\rm X}\,\mu}_{\rm a} +\Psi k_{\rm a}^{\,\mu}
\, .\label{112}\fe
It was already made apparent by (\ref{103a}) in the preceeding section that
the combination on the left of this equation will be conserved under rather
general circumstances: all that is required is that the system as a whole
should be effectively isolated from external influences (other than
gravity) and that the generator $k_{\rm a}^{\,\mu}$ should be a
Killing-Cartan vector in the sense specified above.  What has been shown
in the present section is that if we make the more restrictive postulate
that the material system itself is invariant, in the gauge adjusted
sense characterised by (\ref{94}), under the action generated by the
Killing-Cartan vector, and if we also postulate that the separate force
density contributions $f^{_{\rm X}}_{\,\mu}$ (and hence also the separate
decay rates ${\cal D}_{_{\rm X}}$) all vanish individually, then (\ref{111})
will be applicable so as to provide the much stronger conclusion that
{\it each} of the distinct contributions in the sum on the right of
(\ref{112}) will be {\it separately} conserved.

\bigskip
{\bf 6. Generalised Joukowski theorem}
\medskip

As an application of the generalised Bernouilli theorem discussed in the previous section, the Joukowski formula~\cite{Landau}
is extended to the case of multiconstituent fluid. We shall derive the Magnus
force arising on a perturbing vortex moving in an {\it asymptotically uniform} medium characterised by vanishing currents
$\overline{n_{_{\rm X}}}^{\,\nu} = 0$ (asymptotic values will be denoted by an overhead bar). The gravitational potential is
supposed to be unaffected by the perturbation.
We shall follow closely a previous analysis in a relativistic context~\cite{CarterLangloisPrix00}. The fluid flow
is supposed to be stationary thereby admitting the ether vector $e^{\mu}$ as a Cartan Killing vector, and longitudinally invariant along the
uniform space like Cartan Killing vector $l^{\mu}$ (vortex symmetry axis) satisfying $l^{\mu}t_{\mu}=0$.

The multiconstituent flow is described by the conservation law (\ref{86d})
$ \nabla_{\!\mu}T_{_{\!\rm tot}\nu}^{\ \mu}=0$. Besides each of the currents are separetely conserved
$\nabla_{\!\nu}n_{_{\rm X}}^{\,\nu}=0$. It is further assumed that not only
the total force density vanishes, but that each fluid component is isolated $f_\mu^{_{\rm X}} = n_{_{\rm X}}^{\,\nu}
\varpi^{_{\rm X}}{_{\!\nu\mu}}=0$. It thus
implies that the generalised vorticity 2-form $\varpi^{_{\rm X}}{_{\!\nu\mu}}$ will vanish whenever it is initially zero $\varpi^{_{\rm X}}{_{\!\nu\mu}}=0$.
The framework in which the Joukowski formula is derived, is restricted to such {\it irrotational} flows (whether it is superfluid ot not).

Consequently, from (\ref{91d}), the Bernouilli scalars associated with the corresponding Killing vectors are uniform  $\nabla_{\!\nu}{\cal B}^{_{\rm X}}_{\rm a}= 0$.
In particular, the following Bernouilli scalars are constants:
\be{\cal B}^{_{\rm X}}_{\rm 0}=\pi^{_{\rm X}}_{\,\nu} e^{\nu} = \overline{\pi^{_{\rm X}}}_{\!\nu} e^{\nu}\fe
\be{\cal B}^{_{\rm X}}_{\rm -1}=\pi^{_{\rm X}}_{\,\nu} l^{\nu} = \overline{\pi^{_{\rm X}}}_{\!\nu} l^{\nu}\, .\fe

The multiconstituent fluid is exerting a force (per unit length) on the vortex given by
 \be {\cal F}_{\nu} = \oint T_{_{\!\rm tot}\nu}^{\ \sigma}\, ^{\star}\varepsilon_{\sigma \mu}dx^{\mu}\, ,\fe in which the ``background''
 antisymmetric measure
 tensor is
 defined by \be ^{\star}\varepsilon_{\mu \nu} = \varepsilon_{\mu \nu\rho\sigma}e^{\rho}l^{\sigma}\, .\fe

 The conservation of the total stress energy momentum tensor allows one to evaluate the integral along any closed circuit, which for convenience
 can be chosen sufficently far away from the vortex core for a linear expansion to be valid:
\be T_{_{\!\rm tot}\nu}^{\ \sigma} = \overline{ T_{_{\!\rm tot}}}_{\nu}^{\! \sigma} + \delta  T_{_{\!\rm tot}\nu}^{\ \sigma}+O\bigl(\delta^2\bigr)\, .\fe The
 force is zero by definition in the unperturbed uniform medium hence it reduces to \be {\cal F}_{\nu} = \delta {\cal F}_{\nu}+O\bigl(\delta^2\bigr)\, .\fe

The linear deviation of the stress energy momentum tensor from the uniform background value is expressible as
\be \delta T_{_{\!\rm tot}\nu}^{\ \sigma} = \ssum \bigl(\overline{\pi^{_{\rm X}}}_{\!\nu}\, \delta n_{_{\rm X}}^{\,\sigma} +
\overline{n_{_{\rm X}}}^{\,\sigma}\, \delta  \pi^{_{\rm X}}_{\,\nu} - \overline{n_{_{\rm X}}}^{\,\rho}\,\delta \pi^{_{\rm X}}_{\,\rho}\, \delta_{\nu}^{\sigma}\bigr)\,. \fe
The force per unit length to lowest order is therefore given by $\delta{\cal F}_{\nu}  = \oint  \delta T_{_{\!\rm tot}\nu}^{\ \sigma}\, ^{\star}\varepsilon_{\sigma \mu}dx^{\mu}\, ,$ i.e.
\be\delta{\cal F}_{\nu}  = \ssum \biggl(\overline{\pi^{_{\rm X}}}_{\!\nu}\oint\delta n_{_{\rm X}}^{\,\sigma}\, ^{\star}\varepsilon_{\sigma\mu}dx^{\mu}
+2 \overline{n_{_{\rm X}}}^{\,\sigma}\,\oint \delta\pi^{_{\rm X}}{_{\,[\nu}}\, ^{\star}\varepsilon{\!_{\sigma]\mu}}dx^{\mu}\biggr)\, .\fe

It is then worthwhile to notice as a consequence of the Bernouilli theorem that the variation of the momentum is purely orthogonal to the vortex
\be \perp^{\sigma}_{\nu}\, \delta\pi^{_{\rm X}}_{\,\sigma} = \delta \pi^{_{\rm X}}_{\,\nu}\, ,\fe where $\perp_{\nu}^{\sigma}$ is the operator of
projection orthogonal to the vortex (i.e. orthogonal to the Killing vectors). The spacetime metric $\delta^{\sigma}_{\nu}$ is decomposible into the sum of the operators
of projection parallel
$\eta^{\sigma}_{\nu}$ and orthogonal $\perp^{\sigma}_{\nu}$ to the vortex. The contravariant antisymmetric measure tensor $^{\star}\varepsilon^{\mu \nu}$ is introduced by
\be ^{\star}\varepsilon^{\mu \nu} = t_{\rho}l_{\sigma}\varepsilon^{\mu\nu \rho\sigma}\, .\fe The covariant component $l_{\sigma}$ of the Killing vector
is well defined since this vector is spacelike. The projection operators are given by 
\be \eta_{\nu}^{\sigma} = e^{\sigma}t_{\nu} + l^{\sigma}l_{\nu} \fe
\be \perp_{\nu}^{\sigma} = -^{\star}\varepsilon^{\mu \sigma}\,^{\star}\varepsilon_{\mu \nu}\, . \fe
The force per unit length therefore becomes
\be\delta{\cal F}_{\nu}  = \ssum \biggl(\overline{\pi^{_{\rm X}}}_{\,\nu}\oint\delta n_{_{\rm X}}^{\,\sigma}\, ^{\star}\varepsilon_{\sigma\mu}dx^{\mu}
+2 \overline{n_{_{\rm X}}}^{\,\sigma}\,\oint \delta\pi^{_{\rm X}}_{\,\rho}\, \perp^{\rho}_{[\nu}\,^{\star}\varepsilon_{\sigma]\mu}dx^{\mu}\biggr)\, .\fe

Using the following identity, \be \perp^{\rho}{\!_{[\nu}} \, ^{\star}\varepsilon{\!_{\sigma]\mu}}= -^{\star}\varepsilon_{\nu\sigma}\,\perp^{\rho}_{\mu}\, ,\fe
the force per unit length acting on the vortex simplifies to \be \delta {\cal F}_{\nu} = \ssum \bigl( \overline{\pi^{_{\rm X}}}_{\,\nu}\,\delta {\cal D}_{_{\rm X}}
+ \overline{n_{_{\rm X}}}^{\,\sigma}\,^{\star}\varepsilon_{\sigma\nu}\,\delta {\cal C}^{_{\rm X}} \bigr)\, ,\fe
 in which the momentum circulation integral ${\cal C}^{_{\rm X}}$ and the current outflux integral ${\cal D}_{_{\rm X}}$ are defined by
  \be{\cal D}_{_{\rm X}} = \oint n_{_{\rm X}}^{\,\sigma}\, ^{\star}\varepsilon_{\sigma \mu} dx^{\mu}\fe
 \be{\cal C}^{_{\rm X}} = \oint \pi^{_{\rm X}}_{\,\nu}\, dx^{\nu}\, .\fe

 The current and vorticity conservation laws ensure that ${\cal C}^{_{\rm X}}$ and ${\cal D}_{_{\rm X}}$ respectively, do not depend on the integration closed path.
 Therefore the circuit can be chosen to lie at a very large distance from the vortex core so that the force is exactely given by the linear perturbation term.
 In cases for which there is no current creation in the vortex core, the current outflux integral will simply vanish ${\cal D}_{_{\rm X}} = 0$ and since the asymptotic
 values of these two integrals must also be equal to zero the force per unit length acting on the vortex is eventually given by

 \be {\cal F}_{\nu} = ^{\star}\!\varepsilon_{\sigma\nu}\sum{\cal C}^{_{\rm X}}\overline{n_{_{\rm X}}}^{\,\sigma}\,.\fe

\bigskip
{\bf 7. Virial moment theorems for isolated system}
\medskip

The preceeding analysis has been essentially local, but of course
whenever one is concerned with a system - such as a star model - that
is confined within a compact region it is of particular interest to
consider global quantities, particularly those that are subject
to simple evolution equations or actually conserved.

A specially noteworthy example is of course the total energy
$E_{_{\rm tot}}$ which is definable as a function of
time by a space integral that will be expressible in the form
\be E_{_{\rm tot}}=\int U_{_{\rm tot}}\,{\rm d}^3\! X\, ,\label{132}\fe
with respect to standard Aristotelian coordinates $\{X^{_0},X^i\}$ of the
kind described in Section 2 of \cite{CCI}, with time coordinate $X^{_0}=t$, and space
coordinates $X^i$ of ordinary Cartesian type, so that the 3-dimensional
Euclidean metric components $\gamma_{ij}$ are those of a 3 by 3 unit matrix,
while the time covector has components $t_{_0}=1$, $t_i=0$, and the
corresponding ether velocity vector has components $e^{\,_0}=1$, $e^i=0$.

It is evident from the local energy momentum conservation law
(\ref{86d}) that the time evolution of this quantity will be expressible
as the space integral of a divergence in the form
\be {{\rm d}\over{\rm d} t}E_{_{\rm tot}}=\int \nabla_{\! i}
T_{_{{\rm tot}\, 0}}^{\ i} \,{\rm d}^3\! X\, .\label{133a}\fe
Using Green's theorem to convert this to an asymptotic surface integral
at large distance, it can be seen to follow that for an isolated system
the energy will actually be conserved, i.e. we shall simply have
\be {{\rm d}\over{\rm d} t}E_{_{\rm tot}}=0\, ,\label{133}\fe
provided that (as is always possible in such a case) the gauge for the
gravitational potential $\phi$ in the formula (\ref{86}) for the external
 field contribution $T_{_{{\rm grf}\, 0}}^{\ i}$ is chosen in accordance
with the usual convention that it should tend to zero at large distances.
(This condition automatically entails that $\phi$ will fall off as
the inverse of the radial distance, a requirement which is sufficient
to get rid of the boundary term that would otherwise be left over.)

The foregoing total energy conservation law depends only on the time
projected part of the complete local energy momentum conservation law
(\ref{86d}), which can be decomposed in terms of separate energy density
and 3-momentum density components
\be U_{_{\rm tot}}= -T_{_{{\rm tot}\, 0}}^{\ _0}\, ,\hskip 1 cm
{\mit \Pi}^i=T_{_{\rm tot}}^{\ _0 i}=
\gamma^{ik} T_{_{\rm tot}k}^{\ _0}\, ,\label{130}\fe
in the form
 \be \nabla_{\!_0}U_{_{\rm tot}}=\nabla_{\! i}T_{_{{\rm tot}\, 0}}^{\ i}
\, , \label{131a}\fe
(the part from which (\ref{133a}) is derived) and
\be \nabla_{\!_0}{\mit \Pi}^i=-\nabla_{\! j}T_{_{\rm tot}}^{\ ji}
= -\gamma^{ik}\nabla_{\! j}T_{_{\rm tot}k}^{\ j}
\, . \label{131}\fe
(in which, as remarked above there is no need to distinguish between
${\mit \Pi}^i$ and ${\mit\Pi}_{_{\rm tot}}^{\ i}$ since there is no
purely gravitational 3-momentum contribution). The space projected part,
namely the local 3-momentum conservation law (\ref{131}) is particularly
informative when used in conjunction with the conservation law
(I-108) for the Newtonian mass current, whose time and space components
\be \rho^{_0}=\rho\, ,\hskip 1 cm  \rho^i={\mit \Pi}^i \label{121}\fe
must satisfy
\be \nabla_{\!_0}\rho =-\nabla_{\! i}\,\rho^i\, .\label{122}\fe
By combining this with (\ref{131}) we obtain a noteworthy
second order differential relation,
\be \nabla_{\!_0}\nabla_{\!_0}\rho=\nabla_{\! i}\nabla_{\! j}
T_{_{\rm tot}}^{\, ij}\, ,\label{131b}\fe
that may appropriately be referred to as the local Newtonian virial
equation, since, as will be shown below, it is what ultimately accounts
for various older (specialised) and newer (more general) variants of
what is commonly referred to by the term ``virial theorem''.

The relevant global applications of the foregoing local conservation
laws involve global mass - moment integrals of the form
\be {\cal M}=\int\rho\,{\cal X}\, {\rm d}^3\! X\,\label{150}\fe
and momentum - moment integrals of the form
\be {\cal J}=\int{\mit\Pi}^i\,{\cal Y}_i\, {\rm d}^3\! X\,\label{150a}\fe
where ${\cal X}$ and ${\cal Y}_i$ are weighting factors that are constructed
as fixed (i.e. time independent) given functions of the Cartesian
space coordinates $X^i$. In the same way as (\ref{133}) was obtained
by applying  (\ref{131a}) to (\ref{132}) using Green's theorem to
get rid of a divergence, it can be seen that -- for a locally confined
system --- application of (\ref{122}) in (\ref{150}) gives
\be {{\rm d}\over{\rm d} t}{\cal M}=\int {\mit \Pi}^i\nabla_{\!i} {\cal X}
\, {\rm d}^3\!  X \, .\label{151}\fe
In order for application of (\ref{131}) in (\ref{150a}) to provide
the analogous relation
 \be {{\rm d}\over{\rm d} t}{\cal J}=\int T_{_{\rm tot}}^{\, i j}
\nabla_{\!i} {\cal Y}_j \, {\rm d}^3\!  X \, ,\label{151a}\fe
it is not enough to require that the material system be locally confined:
in order for Green's theorem to be able to get rid of the relevant
divergence contribution the weighting factor ${\cal Y}_i$ must satisfy
an appropriate admissibility restriction. Since, according to (\ref{86}),
the components of the gravitational field contribution
$T_{_{\rm grf}}^{\, i j}$ will fall off as the fourth power of the
radial distance from the material source, it can be seen that the
criterion for admissibility in (\ref{151a}) is that the components
${\cal Y}_i$ should grow more slowly than the square of the radial
distance. Subject to the proviso that ${\cal X}$ should satisfy this
admissibility condition when one makes the substitution ${\cal Y}_i
=\nabla_{\! i}{\cal X}$, it can be seen by combining (\ref{151})
and (\ref{151a})-- or directly from the local virial relation
(\ref{131b}) -- that we shall  obtain a generic global virial relation
of the form
\be {{\rm d}^2\over{\rm d} t^2}{\cal M}=\int
T_{_{\rm tot}}^{\ ij}\nabla_{\!i}\nabla_{\!j}{\cal X}\,{\rm d}^3\! X
\, .\label{152}\fe
The simplest example of the application of the foregoing
formulae is of course to the case of the total mass $M$, which is
obtained simply by taking the weighting factor ${\cal X}$ to be
unity. Thus by setting
\be {\cal X}\mapsto 1\hskip 1 cm \Rightarrow \hskip 1 cm
{\cal M}\mapsto M\, ,\hskip 1 cm
\nabla_{\!i}{\cal X}\mapsto 0 \, ,\label{160}\fe
we see that (\ref{150}) and (\ref{151}) reduce respectively to the
definition and conservation law for the total mass as given by
\be  M=\int\rho\, {\rm d}^3\! X\, ,\hskip 1 cm
{{\rm d}\over{\rm d} t}M=0\, .\label{123}\fe

The next simplest possibility is the case of the dipole moment,
which is obtained by taking  the weighting factor to
be linearly dependent on the Cartesian space coordinates $X^i$.
Thus in particular the dipole moment in the direction of let us say
the $X^{_3}$ axis is obtained just by taking the weighting factor
${\cal X}$ to be $X^{_3}$. Thus by setting
\be {\cal X}\mapsto X^{_3}\hskip 1 cm \Rightarrow \hskip 0.6 cm
{\cal M}\mapsto {\cal D}^{_3}\, ,\hskip 0.6 cm
\nabla_{\!i}{\cal X}\mapsto \gamma^{_3}_{\,i}\, , \hskip 0.6 cm
\nabla_{\!i}\nabla_{\!j}{\cal X}\mapsto 0 \, ,\label{163}\fe
we see that (\ref{150}) and (\ref{151}) reduce respectively to the
definition and variation law for the dipole moment component
${\cal D}^{_3}$  as given by
\be  {\cal D}^{_3}=\int\rho\, X^{_3}\, {\rm d}^3\! X\, ,\hskip 1 cm
{{\rm d}\over{\rm d} t}{\cal D}^{_3}=\int {\mit\Pi}^{_3}\,
{\rm d}^3\! X\, ,\label{163a}\fe
where the latter integral is interpretable as the ordinary linear
3-momentum in the direction of the $X^{_3}$ axis, whose conservation
is now given by the relevant application of (\ref{151a}), or
equivalently of (\ref{152}) which can be seen to reduce simply to
\be {{\rm d}^2\over{\rm d} t^2}{\cal D}^{_3}=0\, .\label{164}\fe
This particular application will of course reduce to a triviality
if we exploit the freedom to use the centre of mass frame
characterised by the condition that the dipole moment simply vanishes.

A less trivial application is to the case of angular momentum,
whose components are obtained by taking the weighting factor
${\cal Y}_i$ to be given by the corresponding components of the
Killing - Yano 2-form (\ref{99b}). Thus in particular the angular
momentum $J_{_3}$ about the $X^{_3}$ axis will be obtained by
identifying the weighting components ${\cal Y}_i$ with the
Killing - Yano components $Y_{i_3}=\varepsilon_{i_3 k}X^{_k}$.
Thus by setting
\be {\cal Y}_i\mapsto Y_{i_3}\hskip 1 cm \Rightarrow \hskip 0.6 cm
{\cal J}\mapsto J_{_3}\, ,\hskip 0.6 cm
\nabla_{\!i}{\cal Y}_j\mapsto \varepsilon_{ij_3}\, ,\label{165}\fe
we see that (\ref{150a}) and (\ref{151a}) reduce respectively to the
definition and the conservation law for this angular momentum component,
as given by
\be  J_{_3}=\varepsilon_{ij_3}\int X^i\, {\mit \Pi}^j{\rm d}^3\! X
\, ,\hskip 1 cm {{\rm d}\over{\rm d} t} J_{_3}=0\, ,\label{166}\fe
in which the last step, namely the vanishing of the time derivative,
results from the stress tensor symmetry property
\be T_{_{\rm tot}}^{\ [ij]}=0\, ,\label{134a}\fe
which holds as a consequence of the corresponding property (\ref{84b})
of the material contribution, and of the manifest  symmetry of the
gravitational contribution (\ref{86}). The familiar conclusion that
the total angular momentum of an isolated Newtonian system will be
conserved can alternatively be interpreted as an immediate global
consequence, via Green's theorem, of the relevant particular application
of the general local conservation law (\ref{103c}).

Having seen how mass and angular momentum conservation laws of the usual
kind can immediately be recovered as particularly simple special cases
for which the weighting factor is uniform or linearly dependent on the
Cartesian space coordinates -- so that the term on the right of
(\ref{152}) will drop out -- we now come to the less trivial category
of applications for which the term ``virial theorem'' is most commonly
employed~\cite{ShapiroTeukolsky83}, namely cases for which the
weighting factor ${\cal X}$ has homogeneous quadratic dependence
on the Cartesian space coordinates $X^i$. The simplest such possibility
is the isotropic case obtained by taking ${\cal X}$ proportional to
the square of the radial distance $r$ from the center as given (in a
Cartesian system with origin at the center of mass) by
\be r^2=\gamma_{ij}X^i X^j\, \label{170a}\fe
while another obviously important special case is that for which
${\cal X}$ is taken to be proportional to the square of the distance
$\varpi$ from let us say the $X^{_3}$ axis, as defined by
\be \varpi^2=\big(\gamma_{ij}-\delta^{_3}_{\,i}\delta^{_3}_{\,j}\big)
X^i X^j= (X^{_1})^2+(X^{_2})^2\, .\label{170b}\fe

The isotropic case obtained simply by identifying ${\cal X}$ with $r^2$
is that for which ${\cal M}$ is just the ordinary scalar quadrupole
moment $I$. Thus by setting
\be {\cal X}\mapsto r^2\hskip 1 cm \Rightarrow \hskip 1 cm
{\cal M}\mapsto Q\, ,\hskip 1 cm
\nabla_{\!i}\nabla_{\!j}{\cal X}\mapsto 2\gamma_{ij} \, ,\label{171}\fe
one sees from the generic virial relation (\ref{152}) that the scalar
quadrupole moment
\be Q=\int \rho\, r^2\,{\rm d}^3\! X \, .\label{172}\fe
will satisfy a time evolution equation of the form
\be {{\rm d}^2\over{\rm d} t^2}Q=2\int
T_{_{\rm tot}\, i}^{\ i}\,{\rm d}^3\! X
\, .\label{172a}\fe
It evidently follows that for a stationary (i.e. time independent)
configuration the integral on the right of this equation must vanish.

The axial case obtained  by identifying ${\cal X}$ with $\varpi^2$
is that for which ${\cal M}$ is just the moment of inertia
moment $I_{_3}$ about the $X^{_3}$ axis. Thus by setting
\be {\cal X}\mapsto \varpi^2\hskip 1 cm \Rightarrow \hskip 1 cm
{\cal M}\mapsto I_{_3}\, ,\hskip 1 cm
\nabla_{\!i}\nabla_{\!j}{\cal X}\mapsto 2 \big(\gamma_{ij}
-\delta^{_3}_{\,i}\delta^{_3}_{\,j}\big)\, ,\label{171b}\fe
one sees from the generic virial relation (\ref{152}) that this
moment of inertia
\be I_{_3}=\int \rho\, \varpi^2\,{\rm d}^3\! X \, ,\label{173}\fe
will satisfy a time evolution equation of the form
\be {{\rm d}^2\over{\rm d} t^2}I_{_3}=2\int \big(
T_{_{\rm tot}}^{\ _{11}}+T_{_{\rm tot}}^{\ _{22}}\big)
\,{\rm d}^3\! X \, .\label{173a}\fe
It evidently follows again that for a stationary (i.e. time independent)
configuration the integral on the right of this equation must vanish.

The preceeding special isotropic and axial cases can be considered as
combinations of the separate components of the corresponding tensorial
theorem -- as obtained by substituting $X^i X^j$ in place of ${\cal X}$
in (\ref{151}) and (\ref{152}) -- according to which the first and
second time derivatives of the generic quadrupole moment component
\be Q^{ij}=\int \rho\, X^i X^j\,{\rm d}^3\! X \, ,\label{174}\fe
will satisfy  equations of the form
\be {{\rm d}\over{\rm d} t}Q^{ij}=2\int {\mit \Pi}^{(i} X^{j)}
\, {\rm d}^3\!  X \, .\label{127}\fe
and
\be {{\rm d}^2\over{\rm d} t^2}Q^{ij}=2\int T_{_{\rm tot}}^{\ ij}
\,{\rm d}^3\! X \, .\label{174a}\fe
As well as the stationary case, for which the integrals on the right of
(\ref{127}) and (\ref{174a}) must vanish, the time dependent case is also
of particular interest because it is the third time derivative of the
trace free part of the quadrupole moment tensor that provides the source
of gravitational radiation in the Newtonian (weak field, low velocity)
limit of general relativity.

The foregoing conclusions, and in particular the scalar quadrupole
evolution equation (\ref{172a}), can be construed as a concise general
statement of what is commonly referred to --  in more detailed
applications to particular well known cases -- as the ``virial theorem''.
The most widely familiar version~\cite{ShapiroTeukolsky83}
applies just to the case of a single constituent perfect fluid, but
in a recent study of ellipsoidal configurations
Sedrakian and Wasserman~\cite{SedraWasser01} have provided a version
applying to cases in which there is a pair of independently moving fluid
constituents. Our present -- formally very simple -- result (\ref{172a})
actually goes further, not just by allowing for the possibility of more
than two constituents, but also, less trivially, by including allowance
for the effect of entrainment whereby the strong coupling between the
constituents modifies the momenta of the constituent particles
whose effective masses may deviate substantially from their ordinary
rest masses.

To relate our concise new general purpose ``virial theorem'' (\ref{172a})
to the more specialised results that are already well known, it is
instructive to consider the distinct contributions that are involved. To
start with it can be seen from (\ref{87c}) that the total energy
(\ref{132}) will be expressible in the form
\be E_{_{\rm tot}}=E_{_{\rm mat}}+E_{_{\rm grt}}\, ,\label{138}\fe
where the purely material contribution is defined by
\be E_{_{\rm mat}}=\int U_{_{\rm mat}}\,{\rm d}^3X\, ,\label{139}\fe
and the total gravitational binding energy contribution is defined by
\be E_{_{\rm grt}}=E_{_{\rm grf}}+E_{_{\rm pot}}=\int U_{_{\rm grt}}
\,{\rm d}^3X\, ,\label{140}\fe
in which
\be E_{_{\rm grf}}=\int U_{_{\rm grf}}\,{\rm d}^3\! X=
{1\over 8 \pi {\rm G}}\int g^i g_i\,{\rm d}^3\! X \, ,\label{140a}\fe
and
\be E_{_{\rm pot}}=\int U_{_{\rm pot}}\,{\rm d}^3\! X=
\int\phi\rho\,{\rm d}^3X\, .\label{140b}\fe
For a confined source, it can be seen from (\ref{88b}) using Green's
theorem that the total gravitational contribution (\ref{140}) will be related
to the separate gravitational field contribution (\ref{140a}) and
gravitational potential contribution (\ref{140b}) by
\be E_{_{\rm grt}} =-E_{_{\rm grf}}={_1\over^2}E_{_{\rm pot}}
\, .\label{141}\fe
In a similar manner, using the observation that
\be T_{_{\rm grt}i}^{\ i}=-U_{_{\rm grf}}\, ,
\label{142a}\fe
it can be seen that the integral on the right hand side of (\ref{152})
can be split as a sum of a purely material contribution and
a gravitational contribution, in which (again using Green's theorem)
the latter works out to be the same -- for  a confined source -- as the
corresponding contribution to the energy, i.e. one obtains
\be
\int T_{_{\rm tot}i}^{\ i}\,{\rm d}^3\!X=
\int T_{_{\rm mat}i}^{\ i}\,{\rm d}^3\!X+ E_{_{\rm grt}}\, .\label{142}\fe

The material energy contribution can evidently be further decomposed as
\be E_{_{\rm mat}}=E_{_{\rm kin}}+E_{_{\rm int}}\, ,\label{143}\fe
where the separate kinetic and internal contributions are expressible
in terms of the densities given by (\ref{83h}) as
\be E_{_{\rm kin}}=\int U_{_{\rm kin}}\,{\rm d}^3\! X\, ,\hskip 1 cm
E_{_{\rm int}}=\int U_{_{\rm int}}\,{\rm d}^3\! X\, .\label{144}\fe
of  which the latter will be expressible in terms of the
rest frame chemical potentials $\chi^{_{\rm X}}$ introduced in
(\ref{83e}) as
\be E_{_{\rm int}}=\ssum \int n_{_{\rm X}}\chi^{_{\rm X}}\,{\rm d}^3\! X
-\int \Psi \,{\rm d}^3\! X\, .\label{144a}\fe

The corresponding decomposition for the material contribution in (\ref{142a})
has the form
\be \int T_{_{\rm mat}i}^{\ i}\,{\rm d}^3\!X= \int T_{_{\rm kin}i}^{\ i}
\,{\rm d}^3\!X+\int T_{_{\rm int}i}^{\ i}\,{\rm d}^3\!X\, ,\label{145}\fe
in which the kinetic contribution can be seen from (\ref{83k}) and
(\ref{83h}) to be given by
\be \int T_{_{\rm kin}i}^{\ i}\,{\rm d}^3\!X=2 E_{_{\rm kin}}
\, ,\label{146}\fe
while the internal contribution can be seen from  (\ref{83a}) to
be given by
\be \int T_{_{\rm int}i}^{\ i}\,{\rm d}^3\!X=3\int\Psi\,{\rm d}^3\!X
\, + \ssum \int n_{_{\rm X}}^{\,i} \chi^{_{\rm X}}_{\,i}\,{\rm d}^3\!X
\, .\label{147}\fe

Putting the separate pieces together, it can be seen that the integral in the
formulation (\ref{172a}) of the generalised ``scalar virial theorem'' --
and that  in any stationary equilibrium configuration must therefore vanish, i.e.
\be \int T_{_{\rm tot}i}^{\ i}\,{\rm d}^3\! X=0\, ,\label{136b}\fe
 -- will be expressible by
\be \int T_{_{\rm tot}i}^{\ i}\,{\rm d}^3\! X
=E_{_{\rm grt}}+2E_{_{\rm kin}}
+3\int\Psi\,{\rm d}^3\!X+\ssum \int n_{_{\rm X}}^{\,i} \chi^{_{\rm X}}_{\,i}
\,{\rm d}^3\!X \, .\label{148}\fe
In the combination on the right hand side -- which must vanish for a
stationary equilibrium configuration -- it is to be remarked that
the first two terms (the total gravitational energy plus
twice the ordinary kinetic energy) are of the kind that is
traditionally familiar. As compared with the multiconstituent version
given recently by Sedrakian and Wasserman~\cite{SedraWasser01}
for the idealised  case in which mutual entrainment between the
constituents was ignored, the difference here consists firstly
of the use of the undecomposable pressure function $\Psi$ in
place of a sum (of the form $\ssum P_{_{\rm X}}$) of ordinary pressure
contributions from the distinct constituents, and secondly of the
inclusion of the final term involving the previously
ignored entrainment momenta $\chi^{_{\rm X}}_{\,i}$ themselves.

Before concluding, we wish to point out that the homogeneously
quadratic category that has been discussed in detail immediately
above is not the only useful category of non-trivial applications
of our generic virial theorem (\ref{152}). Another category worth
consideration is that of homogeneously first (as opposed to second)
order mass moments. Strictly linear dependence on the Cartesian space
coordinates leads merely to the trivial dipole moment case discussed
above, but a no less natural, and much more interesting, alternative
possibility is that  of the isotropic homogeneously first order (but
non-linear) case obtained by setting ${\cal X}=r$, while the next most
obviously interesting possibility is that of  the cylindrical
homogeneously first order (but non linear)  case obtained by setting
${\cal X}=\varpi$.

For the first of these, namely the isotropic first order case, we have
\be {\cal X}=r\hskip 1 cm \Rightarrow\hskip 1 cm
\nabla_{\!i}\nabla_{\!j}{\cal X}=r^{-1}
(\gamma_{ij}-\nu_{r i} \nu_{r j})\, ,\label{155}\fe
where $\nu_r^{\,i}$ is the radial unit vector as defined by
\be \nu_r^{\,i}=r^{-1} X^i\, .\label{155a}\fe
Since, in terms of standard spherical coordinates $r,\theta,\varphi$,
the volume element will be given by ${\rm d}X^{_1}\,{\rm d}X^{_2}\,
{\rm d}X^{_3}= r^2\,{\rm sin}\,\theta\,{\rm d}r\, {\rm d}\theta\,
{\rm d}\varphi$, the isotropic homogeneously linear virial theorem
obtained by substituting (\ref{155}) in (\ref{152}) will be expressible as
\be {{\rm d}^2\over{\rm d} t^2}\int \rho\, r^3\,{\rm sin}\,\theta\,
{\rm d}r\, {\rm d}\theta\, {\rm d}\varphi=\int \big(T_{_{\rm tot}i}^{\ i}
-T_{_{\rm tot}}^{\ ij}\nu_{r\, i} \nu_{r\, j}\big)r\,{\rm sin}\,
\theta\,{\rm d}r\, {\rm d}\theta\, {\rm d}\varphi\, .\label{155b}\fe

An even simpler relation is obtainable for the cylindrical
first order case, for which we have
\be {\cal X}=\varpi\hskip 1 cm \Rightarrow\hskip 1 cm
\nabla_{\!i}\nabla_{\!j}{\cal X}=\varpi^{-1}\nu_{\varphi\, i}
\nu_{\varphi\, j} \, ,\label{156}\fe
where $\nu_\varphi^{\,i}$ is the unit vector in the direction of the
relevant axial Killing vector (\ref{99a}), namely
\be \nu_\varphi^{\,i}=\varpi^{-1} k_3^{\,i} \,  ,\label{156a}\fe
so that its components will be given by $ \nu_\varphi^{\, _1}=
-X^{_2}/\varpi$, $ \nu_\varphi^{\, _2}=X^{_1}/\varpi$,
$\nu_\varphi^{\, _3}=0$. This leads to what is interpretable as the
Newtonian limit~\cite{GourgBona93} of a result originally derived in a
relativistic context by Bonazzola~\cite{Bona73}. In terms of standard
cylindrical coordinates, as specified by $X^{_1}=\varpi\,{\rm cos}\,
\varphi$,  $X^{_2}=\varpi\,{\rm sin}\,\varphi$, $X^{_3}=z$, the volume
element will be given by ${\rm d}X^{_1}\,{\rm d}X^{_2}\, {\rm d}X^{_3}=
\varpi\, {\rm d}\varpi\, {\rm d}z\, {\rm d}\varphi$. It can thus be seen
that the cylindrical homogeneously linear virial theorem obtained by
substituting (\ref{156}) in the generic relation (\ref{152}) will take
the form
\be {{\rm d}^2\over{\rm d} t^2}\int \rho\, \varpi^2\, {\rm d}\varpi\,
{\rm d} z\, {\rm d}\varphi=\int T_{_{\rm tot}}^{\ ij}
\nu_{\varphi\,i} \nu_{\varphi\, j}\,{\rm d}\varpi\, {\rm d} z\, {\rm d}
\varphi\, .\label{158}\fe
More particularly, in the stationary axisymmetric case, for which the left
hand side vanishes and for which the $\varphi$ integration is trivial,
(\ref{158}) will reduce to the 2-dimensional integral relation
\be \int T_{_{\rm tot}}^{\ ij}\nu_{\varphi\, i}\nu_{\varphi\, j}
\,{\rm d}\varpi\, {\rm d} z\, =0\,  ,\label{159}\fe
that was originally derived by Gourgoulhon and Bonazzola~\cite{GourgBona93},
in which it can be seen from (\ref{86}) that the gravitational contribution 
to the integrand can be separated out in a decomposition that takes the 
form
 \be T_{_{\rm tot}}^{\ ij}\nu_{\varphi\, i}\nu_{\varphi\, j}
=T_{ij}\,\nu_{\varphi}^{\, i}\nu_{\varphi}^{\, j}-U_{_{\rm grf}}
\, ,\label{159b}\fe
as a consequence of the axisymmetry.

We conclude by emphasising that, like the ordinary virial theorem 
(\ref{136b}), the Bonazzola theorem (\ref{159}) is a particular 
example of the condition 
\be \int T_{_{\rm tot}}^{\ ij}\nabla_{\!i}\nabla_{\!j}{\cal X}
\,{\rm d}^3\! X=0\, ,\label{180}\fe
that can be seen from (\ref{152}) to be a generic requirement for any
stationary equilibrium configuration for all admissible choices of the 
weighting factor ${\cal X}$. It is to be recalled that 
$T_{_{\rm tot}}^{\, ij}$ is the total of all material and gravitational 
stress contributions, and that, for the purpose of this generic theorem, 
the relevant admissibility criterion is  that ${\cal X}$ should be any 
given function of the Cartesian space coordinates whose gradient
components $\nabla_{\! i}{\cal X}$ increase more slowly than a 
quadratic function of these coordinates at large distance. 
(This condition would fail for a substitution of the form 
${\cal X}\mapsto r^3$, but it is evidently satisfied by the substitution
${\cal X}\mapsto r^2$ that leads to the ordinary virial theorem, and
a fortiori by the substitution ${\cal X}\mapsto\varpi$ that leads
to the Bonazzola virial theorem.) Whereas the individual terms in the 
expansion (\ref{147}) have forms that depend on the specific nature 
of the multiconstituent fluid models developped above, on the other hand
the generic relation (\ref{180}) and its dynamical generalisation
(\ref{152}) depend only on the generic form of the energy momentum and 
mass conservation laws (\ref{131a}), (\ref{131}), and (\ref{122}),
which should hold for any (complete) Newtonian continuum model.  These
generally applicable laws are all that is needed for the derivation of 
the local virial equation (\ref{131b}) that underlies the global relation
(\ref{152}). The generic virial equilibrium relation (\ref{180}) is
just the global consequence of the local equilibrium condition
\be \nabla_{\! i}\nabla_{\! j}
T_{_{\rm tot}}^{\, ij}=0\, ,\label{182}\fe
that obviously holds as the stationary specialisation of the local 
virial equation, namely the dynamical relation (\ref{131b}),  whose 
importance -- as an easily memorable law that must be satisfied 
by any Newtonian continuum configuration -- has not been adequately 
recognised hitherto.

\bigskip
{\bf Acknowledgements}
\medskip

The authors wish to thank Silvano Bonazzola, Eric Gourgoulhon, David
Langlois, and Reinhard Prix for instructive conversations.

\bigskip
{\bf Appendix: Noether identities in Newtonian theory}
\medskip

Let us consider the generic case of a Newtonian model characterised by 
a total action density that is a sum
\be \Lambda_{_{\rm tot}}=\sum_a \Lambda_a\, ,\label{A1}\fe
of contributions labelled by some index $a$. This includes the kind of 
model dealt with in the preceeding work which can be described by taking 
$a$ to range over the four values specified by the labels ``kin'', ``int'', 
``pot'' and ``gra''. In what follows we shall use the symbol $\cong$ to 
denote equivalence modulo a space time divergence. The content of the 
action principle is thereby expressible as the postulate that when the
relevant field equations are satisfied, any ``admissible'' infinitesimal 
variation of the relevant dynamical field variables will give rise to a 
corresponding total variation $\delta \Lambda_{_{\rm tot}}$ that is 
equivalent to zero in this sense, i.e. $\delta \Lambda_{_{\rm tot}}
\cong 0$, so that its space time integral will vanish by Green's theorem
for any variation with compact support.

The purpose of the present section is to consider the effects
of more general -- not necessarily ``admissible'' -- variations
of the relevant dynamical fields as well as variations of the
various spacetime background fields on which the complete specification
of the action depends. For such more general variations, the total
action density variation will satisfy a relation of the form
\be \delta\Lambda_{_{\rm tot}}\cong\sum_a\delta^\ddagger \Lambda_a
+\sum_a\delta^\sharp \Lambda_a\, ,\label{A2}\fe
in which $\delta^\ddagger$ denotes the contribution from the variations
of the background fields and $\delta^\sharp$ denotes the contribution
from any non-admissible parts of the variations of the dynamical fields.

In the technically simpler case of a relativistic model the relevant 
spacetime background variation would be fully determined by the variation 
of the space time metric $g_{\mu\nu}$, or equivalently of its contravariant
inverse $g^{\mu\nu}$. However, in the Newtonian case under consideration
here, it is not sufficient to know the variation of the corresponding 
contravariant space metric $\gamma^{\mu\nu}$. It will be necessary as 
well to know the associated variation of the corresponding preferred 
time gradient $t_\mu$. To deal with more elaborate models, the variation 
of the linear connection might also be involved, but this will not be 
necessary for models of the simple kind considered here. However to deal 
with the Galilean gauge dependent  contribution $\Lambda_{_{\rm kin}}$ it
will  also be necessary to take into account the variation of the chosen 
ether frame vector $e^{\mu}$. Thus the relevant background field variations
will be given by expressions of the form
\be \delta^\ddagger\Lambda_a={\partial\Lambda_a\over\partial\gamma^{\mu\nu}}
\,\delta\gamma^{\mu\nu}+{\partial\Lambda_a\over\partial t_\mu}\,\delta t_\mu
+{\partial\Lambda_a\over \delta e^\mu}\,\delta e^\mu\, ,\label{A3}\fe
in which the final term will drop out for gauge independent contributions
such as $\Lambda_{_{\rm mat}}$, $\Lambda_{_{\rm pot}}$, and
$\Lambda_{_{\rm grf}}$. In the models under consideration, the relevant
dynamical fields are the gravitational potential, $\phi$, for which
arbitrary variations are admissible, and the current vectors
$n_{_{\rm X}}^{\,\mu}$ whose admissible variations are restricted,
so that a generic variation $\delta n_{_{\rm X}}^{\,\mu}$ may include
a non - admissible part $\delta^\sharp n_{_{\rm X}}^{\,\mu}$ that
provides a contribution of the form
\be \delta^\sharp \Lambda_a={\partial \Lambda_a\over\partial
 n_{_{\rm X}}^{\,\mu}}\,\delta^\sharp n_{_{\rm X}}^{\,\mu}
\, .\label{A4}\fe

As in the (for this purpose simpler) relativistic case, a standard
procedure~\cite{Trautman65} for the derivation of useful Noether type
identities is to consider variations of the trivial kind generated by an
arbitrary displacement vector field, $\xi^\mu$ say. This means that the
variation of each (background or dynamical) field variable will be given by
the negative of its Lie derivative. The relevant formulae are thus given by
\be -\delta\Lambda_a={\vec \xi}\Libra\,\Lambda_a\equiv\xi^\rho\nabla_{\!\rho}
\Lambda_a,\, \label{A5a}\fe
\be -\delta\gamma^{\mu\nu}={\vec\xi}\Libra\,\gamma^{\mu\nu}\equiv
\xi^\rho\nabla_{\!\rho}\gamma^{\mu\nu}-2\gamma^{\rho\,(\mu}
\nabla_{\!\rho}\xi^{\nu)}\, ,\label{A5b}\fe
\be -\delta t_\mu={\vec\xi}\Libra\, t_\mu\equiv
\xi^\rho\nabla_{\!\rho}t_\mu+t_\rho\nabla_{\!\mu}\xi^\rho\, ,\label{A5c}\fe
\be -\delta e^\mu={\vec\xi}\Libra\, e^{\mu}\equiv
\xi^\rho\nabla_{\!\rho}e^\mu-e^\rho\nabla_{\!\rho}\xi^\mu\, ,\label{A5d}\fe
and
 \be -\delta  n_{_{\rm X}}^{\,\mu}={\vec\xi}\Libra\, n_{_{\rm X}}^{\,\mu}
\equiv \xi^\rho\nabla_{\!\rho} n_{_{\rm X}}^{\,\mu} -n_{_{\rm X}}^{\,\rho}
\nabla_{\!\rho}\xi^\mu\, .\label{A6}\fe
In view of the uniformity properties (I-16) and (I-18)
of the unperturbed background fields, we shall be left with
\be \delta\gamma^{\mu\nu}=2\gamma^{\rho\,(\mu}
\nabla_{\!\rho}\xi^{\nu)}\, ,\label{A7a}\fe
\be \delta t_\mu=-t_\rho\nabla_{\!\mu}\xi^\rho\, ,\label{A7b}\fe
\be \delta e^\mu=e^\rho\nabla_{\!\rho}\xi^\mu\, ,\label{A7c}\fe
and it can be seen from the expression (I-86) for the form
of an admissible current variation that the only remaining
(i.e. inadmissible) contribution
from (\ref{A6}) will be given simply by
\be\delta^\sharp n_{_{\rm X}}^{\,\mu}=n_{_{\rm X}}^{\,\mu}
\nabla_{\!\rho} \xi^\rho\, .\label{A8}\fe

Since (using the notation introduced in (\ref{A2}) for equivalence modulo a
divergence) the variation (\ref{A5a}) will evidently satisfy
\be \delta\Lambda_a\cong\Lambda_a \nabla_{\!\rho}\xi^\rho\, \label{A9}\fe
it can be seen that the terms in (\ref{A2}) can be regrouped on the
left to give a relation of the simple form
\be T_{_{\!\rm tot}\,\nu}^{\ \mu}\nabla_{\!\mu}\xi^\nu\cong 0
\, ,\label{A10}\fe
in which the quantity that will be interpretable as the relevant
total stress momentum energy density tensor can be read out in the
form
\be T_{_{\!\rm tot}\,\nu}^{\ \mu}=\sum_a T_{a\,\nu}^{\, \mu}
\, ,\label{A11}\fe
as the sum of contributions given by the formula
\be T_{a\,\nu}^{\, \mu}=\Psi_a\delta^\mu_{\,\nu}-
2{\partial\Lambda_a\over\partial\gamma^{\rho\nu}}\,\gamma^{\rho\mu}
+ {\partial\Lambda_a\over \partial t_\mu}\, t_\nu -{\partial
\Lambda_a\over\partial e^\nu}\, e^\mu\, ,\label{A12}\fe
in which the relevant generalised pressure contribution $\Psi_a$
is given in terms of the corresponding momentum contributions,
\be \pi_{ a\ \mu}^{\,_{\rm X}}={\partial \Lambda_a\over
\partial n_{_{\rm X}}^{\,\mu}}\, ,\label{A13}\fe
by
\be \Psi_a =\Lambda_a -\ssum \pi_{ a\ \mu}^{\,_{\rm X}}
n_{_{\rm X}}^{\,\mu}\, .\label{A14}\fe

It is to be remarked that, due to the restrictions
\be t_\nu\,\delta \gamma^{\nu\mu}=-\gamma^{\mu\nu}\,\delta t_\nu
\, ,\hskip 1 cm t_\nu\,\delta e^\nu=-e^\nu\,\delta t_\nu\, ,
\label{A15}\fe
resulting from (I-3) and (I-16), there is some gauge ambiguity
in the specification of the partial derivative coefficients introduced in
(\ref{A3}) but it can easily be checked that the stress momentum energy
contributions $T_{a\,\nu}^{\, \mu}$ specified by the combination (\ref{A12})
will be physically well defined in the sense of being unaffected by the
choice adopted in (\ref{A2}).

By a further equivalence transformation (modulo a divergence) it can be
seen that (\ref{A10}) can be converted to the form
\be \xi^\nu\nabla_{\!\mu}T_{_{\!\rm tot}\,\nu}^{\ \mu}\cong 0
\, .\label{A16}\fe
Since this must hold for an arbitrary vector field $\xi^\mu$, and hence
in particular for a displacement field with compact support in any small
space-time neighbourhood, it can be seen by integrating over such a
neighbourhood (so that the divergence ambiguity in the equivalence
relation cancels out by Green's theorem,  as in the usual derivation of
the dynamical equations from the action principle) that the coefficient of
$\xi^\mu$ in (\ref{A16}) must vanish, i.e. we obtain a total energy
momentum conservation law of the form
\be \nabla_{\!\mu}T_{_{\!\rm tot}\,\nu}^{\ \mu}=0\, .\label{A17}\fe

It remains to show that the conserved total stress momentum energy
density tensor $T_{_{\!\rm tot}\,\nu}^{\ \mu}$ obtained by the
foregoing procedure is actually the same as the tensor
$T_{_{\!\rm tot}\,\nu}^{\ \mu}$ introduced in (\ref{86a})

The relation (\ref{A17}) is not an identity in the strict sense, since
its derivation depends on the dynamical field equations obtained
from the action principle. To obtain a purely mathematical identity it is
necessary to take account of the unrestrained variations of all relevant
field variables. In the application with which we are concerned here this
includes not just the background field variations involved in (\ref{A3})
and the variations of the currents $n_{_{\rm X}}^{\,\mu}$ but also the
variations of the gravitational potential $\phi$ and its gradient, so that
the complete generic expression for the variation of an action density
contribution $\Lambda_a$ will have the form
\be \delta\Lambda_a=\delta^\ddagger \Lambda_a
+\ssum \pi_{ a\ \mu}^{\,_{\rm X}}\,\delta
n_{_{\rm X}}^{\,\mu}+ {\partial \Lambda_a\over\partial \phi}
\delta\phi+{\partial\Lambda_a\over\partial(\phi_{,\mu})}
\,\delta(\phi_{,\mu})\, .\label{A18}\fe
For variations generated, as before, by an arbitrary displacement vector
field $\xi^\mu$, we can again use the background field variation formulae
(\ref{A7a}), (\ref{A7b})and (\ref{A7c}) together with the corresponding
dynamical field variation formulae (\ref{A6}) and
\be \delta\phi=-\vec\xi\Libra\,\phi=-\xi^\mu\phi_{,\mu}
\, ,\label{A19}\fe
we thus obtain a relation of the form
$$\xi^\nu\Big(\nabla_{\!\nu}\Lambda_a-\ssum_{_{\rm X}}
\pi_{ a\ \mu}^{\,_{\rm X}}\,\nabla_{\!\nu}
n_{_{\rm X}}^{\,\mu}-{\partial \Lambda_a\over\partial \phi}
\nabla_{\!\nu}\phi-{\partial\Lambda_a\over\partial(\phi_{,\mu})}
\,\nabla_{\!\mu}\nabla_{\!\nu}\phi\Big) $$
\be =\Big({\partial\Lambda_a\over\partial(\phi_{,\mu})}\,\nabla_{\!\nu}\phi
-\ssum_{_{\rm X}}\pi_{ a\ \nu}^{\,_{\rm X}}\, n_{_{\rm X}}^{\,\mu}
-2{\partial\Lambda_a\over\partial\gamma^{\rho\nu}}\,\gamma^{\rho\mu}
+ {\partial\Lambda_a\over \partial t_\mu}\, t_\nu -{\partial
\Lambda_a\over\partial e^\nu}\, e^\mu\Big)
\nabla_{\!\mu}\xi^\nu\, ,\label{A20}\fe
which will be satisfied as an identity in the sense of being independent
of the dynamical field equations. Since it is possible to choose both
$\xi^\nu$ and its derivative $\nabla_{\!\mu}\xi^\nu$ independently at any
given point, it follows that the corresponding coefficients must vanish.
Thus from the left hand side of (\ref{A20}) we obtain the obvious identity
\be\nabla_{\!\nu}\Lambda_a=\ssum \pi_{ a\ \mu}^{\,_{\rm X}}
\,\nabla_{\!\nu} n_{_{\rm X}}^{\,\mu}+{\partial \Lambda_a\over\partial
 \phi} \nabla_{\!\nu}\phi+{\partial\Lambda_a\over\partial(\phi_{,\mu})}
\,\nabla_{\!\mu}\nabla_{\!\nu}\phi\, ,\label{A21}\fe
while from the righthand  side we obtain an identity of the less trivial
form
\be {\partial\Lambda_a\over\partial(\phi_{,\mu})}\,\nabla_{\!\nu}\phi
-\ssum \pi_{ a\ \nu}^{\,_{\rm X}}\, n_{_{\rm X}}^{\,\mu}
-2{\partial\Lambda_a\over\partial\gamma^{\rho\nu}}\,\gamma^{\rho\mu}
+ {\partial\Lambda_a\over \partial t_\mu}\, t_\nu -{\partial
\Lambda_a\over\partial e^\nu}\, e^\mu=0\, .\label{A22}\fe

A noteworthy special case is that of the internal contribution
$\Lambda_{_{\rm int}}$, which depends neither on the
gravitational potential $\phi$ (unlike $\Lambda_{_{\rm pot}}$)
nor on the ether frame vector $e^\mu$ (unlike $\Lambda_{_{\rm kin}}$),
so that when it is substituted for $\Lambda_a$ in (\ref{A22})
the first and the last term will both drop out. Contraction with
$t_\mu$ and $\gamma^{\nu\sigma}$ can then be used to eliminate
the other terms at the end, leaving a result that can be recognised
as the simple identity (I-143) that was quoted above, while
the following identity (I-144) can similarly be derived by
performing an antisymmetrisation after contracting just with
$\gamma^{\nu\sigma}$.

Quite generally, the Noether identity (\ref{A22}) can be used to replace 
the formula (\ref{A12}) by the more explicit and manifestly gauge 
independent expression
\be T_{a\,\nu}^{\, \mu}=\Psi_a\delta^\mu_{\,\nu}+
\ssum \pi_{ a\ \nu}^{\,_{\rm X}}\, n_{_{\rm X}}^{\,\mu}
-{\partial\Lambda_a\over\partial(\phi_{,\mu})}\,\nabla_{\!\nu}\phi
\, .\label{A23}\fe
With the aid of (\ref{A21}) it can be directly verified that each such 
contribution will satisfy a divergence identity of the form
\be \nabla_{\!\mu}T_{a\,\nu}^{\, \mu}=\ssum f_{a\, \nu}^{_{\rm X}}
+{\delta\Lambda_a\over\delta \phi}\,\nabla_{\!\nu}\phi\, ,\label{A24}\fe
in which the Eulerian derivative has a definition of the usual form
\be {\delta\Lambda_a\over\delta \phi}={\partial\Lambda_a\over\partial
 \phi}-\nabla_{\!\mu}\Big({\partial\Lambda_a\over\partial(\phi_{,\mu})}
\Big)\, ,\label{A25}\fe
while the force density contributions are given by an expression 
whose form 
\be f_{a\,\mu}^{_{\rm X}} = 2 n_{_{\rm X}}^{\,\nu}\nabla_{\![\nu}
\pi^{_{\rm X}}_a{_{\mu]}}+\pi^{_{\rm X}}_a{_\mu}\nabla_{\!\nu}
n_{_{\rm X}}^{\,\nu}\, .\label{A26}\fe
is analogous to that for the combined force densities (I-159).

By summing over these separate contributions we obtain the
corresponding formula
\be T_{_{\rm tot}\,\nu}^{\ \mu}=\Psi_{_{\rm tot}}\delta^\mu_{\,\nu}+
\ssum_{_{\rm X}}\pi_{_{\rm tot}\, \nu}^{\ _{\rm X}}\, n_{_{\rm X}}^{\,\mu}
-{\partial\Lambda_{_{\rm tot}}\over\partial(\phi_{,\mu})}\,\nabla_{\!\nu}\phi
\, ,\label{A28}\fe
for the total stress momentum energy density tensor, where the
corresponding total pressure function and momenta are given by
\be \Psi_{_{\rm tot}}=\sum_a \Psi_a\, ,\hskip  1 cm
\pi_{_{\rm tot}\, \nu}^{\ _{\rm X}}=\sum_a\pi_{ a\ \nu}^{\,_{\rm X}}
\, ,\label{A29}\fe
while there will be an analogous expression
\be f_{_{\rm tot}\, \nu}^{\ _{\rm X}}=\sum_a f_{ a\ \nu}^{\,_{\rm X}}
\, ,\label{A30}\fe
for the total non-gravitational force density exerted on each constituent.
 
In the application considered in the preceeding work the purely 
gravitational action density $\Lambda_{_{\rm grf}}$ given by (\ref{85a}) 
provides no contribution to the  momenta, i.e. we shall have 
$\pi_{_{\rm grf}\, \nu}^{\ _{\rm X}}=0$, and hence
$\Psi_{_{\rm grf}}=\Lambda_{_{\rm grf}}$ so we can make the 
identifications
$\pi_{_{\rm tot}\, \nu}^{\ _{\rm X}}=\pi_{\, \nu}^{_{\rm X}}$,
$f_{_{\rm tot}\, \nu}^{\ _{\rm X}}=f_{\, \nu}^{_{\rm X}}$,
and $\Psi_{_{\rm tot}}=\Psi+\Lambda_{_{\rm grf}}$
in which it is to be recalled that we use a blank label to
indicate the sum over all values of $a$ except the gravitational
contribution ``gra'', i.e. for the sum over the values
``kin'', ``int'', and ``pot'', which is the only part that is relevant
if (as in the Cowling approximation) one is  concerned just with evolution 
in a fixed gravitational background, but not with the effects of self 
gravity. It can thus be verified that the definitions introduced in the 
systematic mathematical procedure developed in this appendix are entirely 
consistent with those introduced ad hoc in the main part of the text. In 
particular, it can easily be seen from (\ref{A28}) that the general 
purpose prescription provided by (\ref{A12}) and (\ref{A13}) for 
$T_{_{\rm tot}\,\nu}^{\ \mu}$ leads to a result that is in exact agreement 
with what is given by the formula (\ref{86a}) that was obtained from
more specific physical considerations in the main part of the text, while 
similarly by summing over the index $a$ in the identity (\ref{A26}) one 
recovers a result that can be seen to agree with the previously quoted
divergence formula (\ref{81}).


\begin{thebibliography}{99}

\bibitem{CCI} B. Carter, N. Chamel, 
``Covariant analysis of Newtonian multi-fluid models for neutron stars:
I Milne-Cartan structure and variational formulation'', (2003), to be 
published in {\it Int. J. Mod. Phys.} [astro-ph/0305186].

\bibitem{Carter89} B. Carter,
``Covariant Theory of Conductivity in Ideal Fluid or Solid Media",
in {\it Relativistic Fluid Dynamics (C.I.M.E., Noto, May 1987)}
ed.  A.M. Anile, \& Y. Choquet-Bruhat, Lecture Notes in Mathematics {\bf 1385}
(Springer - Verlag, Heidelberg, 1989) 1-64.

\bibitem{CarterKhalatnikov94} B. Carter \& I.M. Khalatnikov,
``Canonically covariant formulation of Landau's Newtonian superfluid
dynamics", 
{\it Rev. Math. Phys.} {\bf 6} (1994)  277-304.


\bibitem{Bonnor57} W.B. Bonnor,
{\it Mon. Not. R.A.S.} {\bf 117} (1957) 104.

\bibitem{Yano55} K. Yano, {\it The theory of Lie derivatives and
its applications} (North Holland, Amsterdam, 1995) 9.

\bibitem{CarterMcLenaghan79} B. Carter, R.G. McLenaghan, 
``Generalised total angular momentum operator for the Dirac equation
in curved space time'', {\it Phys. Rev.} {\bf D19} (1979) 1093-1097.

\bibitem{Landau} L. Landau, E. Lifchitz,
{\it Physique th\'eorique, M\'ecanique des fluides} (Editions Mir, 1989)

\bibitem{CarterLangloisPrix00} B. Carter, D. Langlois, R. Prix,
``Relativistic solution of Iordanskii problem in multiconstituent
superfluid mechanics'', in {\it Vortices in unconventional superconductors
and superfluids} (eds. R.P. Huebener, N. Schopohl, G.E.
Volovik, Springer 2002)

\bibitem{ShapiroTeukolsky83} S.L. Shapiro, S.A. Teukolsky,
{\it Black Holes, White Dwarfs, Neutron Stars} (Wiley, New York, 1983).


\bibitem{SedraWasser01} A. Sedrakian, I. Wasserman,
``The tensor virial method and its application to self gravitating
superfluids'', in {\it Physics of neutron star interiors}
({Lecture notes in Physics} {\bf 578})
ed D. Blaschke, N. Glendenning, A. Sedrakian
(Springer, Heidelberg, 2001) 97 - 126.


\bibitem{GourgBona93} E. Gourgoulhon, S. Bonazzola,
``A virial identity applied to stellar models'',
{\it Class. Q. Grav.} {\bf 11} (1993) 1775-1784.

\bibitem{Bona73} S. Bonazzola,
{\it Astroph. J.} {\bf 182} (1973) 335

\bibitem{Trautman65} A. Trautman,
``Invariance properties and conservation laws''
in {\it Brandeis Lectures on General Relativity}
ed. A. Trautman, F.A.E. Pirani, H. Bondi
(Prentice Hall, New Jersey, 1965) 158 - 200.

\end{thebibliography}
\end{document}